\documentclass[aps,prd, onecolumn,superscriptaddress, 11pt]{revtex4} 
\pdfoutput = 1
\usepackage{graphicx}
\usepackage{amssymb, amsmath, amsthm, amsfonts, bm, mathrsfs, wasysym, txfonts, bbm, enumerate}
\usepackage[usenames,dvipsnames]{xcolor}
\usepackage{epsfig}
\usepackage{graphicx}
\usepackage{amsfonts}
\usepackage{amssymb}
\usepackage{bm}
\usepackage{latexsym}
\usepackage{amsmath}
\usepackage{hyperref}
\usepackage{xcolor}
\usepackage{bbold}
\begin{document}

\title{{\large A unified composite model of inflation and dark matter in the Nambu--Jona-Lasinio theory} \vspace{3mm}}

\author{Phongpichit Channuie}
\email[]{channuie@gmail.com}
\affiliation{School of Science, Walailak University, Nakhon Si Thammarat, 80160 Thailand}

\author{Chi Xiong}
\email[]{xiongchi@ntu.edu.sg}
\affiliation{Institute of Advanced Studies \& School of Physical and Mathematical Science, Nanyang Technological University, 639673 Singapore \vspace{5mm}}


\begin{abstract}
In this work, we propose a cosmological scenario inherently based on the effective Nambu--Jona-Lasinio (NJL) model that cosmic inflation and dark matter can be successfully described by a single framework. On the one hand, the scalar channel of the NJL model plays a role of the composite inflaton (CI) and we show that it is viable to achieve successful inflation via a non-minimal coupling to gravity. For model of inflation, we compute the inflationary parameters and confront them with recent Planck 2015 data. We discover that the predictions of the model are in excellent agreement with the Planck analysis. We also present in our model a simple connection of physics from the high scales to low scales via renormalization group equations (RGEs) of the physical parameters and use them to estimate the range of relevant parameters. On the other hand, the pseudoscalar channel can be assigned as a candidate for composite dark matter (CD). For model of dark matter, we couple the pseudoscalar to the Higgs sector of the standard model with the coupling strength $\kappa$ and estimate its thermally-averaged relic abundance. We discover that the CD mass is strongly sensitive to the coupling $\kappa$. We find in case of light CD, $M_{s}<M_{h}/2$, that the required relic abundance is archived for value of its mass $M_{s} \sim 61\,{\rm GeV}$ for $\kappa=0.1$. However, in this case the CD mass can be lighter when the coupling is getting larger. Moreover, in case of heavy CD, $M_{s}> M_{W,\,Z}$ (or $>M_{h}$), the required relic abundance can be satisfied for value of the CD mass $M_{s}\sim 410\,{\rm GeV}$ for $\kappa = 0.5$. In contradiction to the light mass case, however, the CD mass in this case can even be heavier when the coupling is getting larger.
\end{abstract}

\maketitle     
         
\section{Introduction}

The observations convince us that the universe is nowadays dominated by unidentified forms of matter, called Dark Matter (DM), and energy, called Dark Energy (DE). The nature of dark matter conveys one of the unsolved problems in physics and also dark energy is still the greatest cosmic mystery. Weakly interacting massive particles (WIMPs) are so far the leading particle candidate for DM, see \cite{Feng:2010gw} for example. However, many other paradigms,  including superWIMPs, e.g. \cite{Feng:2003uy}, light gravitinos and sterile neutrinos, e.g. \cite{Viel:2005qj}, are still possible to account for DM candidates. Another prominent physics problem is cosmic inflation in which the universe went through a period of extremely rapid expansion. The inflationary paradigms \cite{Starobinsky:1979ty,Starobinsky:1980te,Guth:1980zm,Linde:1981mu,Albrecht:1982wi} were initially proposed to solve important issues, e.g. the magnetic monopoles, the flatness, and the horizon problems, and simultaneously provide the mechanism for generation of density perturbations as seed for the formation of large scale structure in the universe.  Nowadays, an inflationary scenario is well established as an indispensable ingredient of modern cosmology. Its predictions fit very well with various experimental data, e.g. Planck collaboration \cite{Ade:2015lrj}. Traditionally, inflationary models were so far modelled via the introduction of new (elementary) scalar fields, e.g. Higgs inflation \cite{Bezrukov:2007ep,Bezrukov:2008ut}. More interestingly, the authors of, for example, \cite{Clark09,Lerner:2009xg} proposed models in which inflation and dark matter can be described on the same footing.

However, the elementary scalar field in field theories is plagued by the so-called hierarchy problem. This problem is commonly meant that quantum corrections generate unprotected quadratic divergences which must be fine-tuned away if the models must be true till the Planck scale. One of the compelling scenarios to solve/avoid the hierarchy problem involves a composite field of some strongly coupled theories, e.g. technicolor, featuring only fermionic matter, and therefore stable with respect to quantum corrections. Therefore, in order to alternatively describe nature, one can imagine that the scalar fields, e.g. the inflaton and dark matter, need not be an elementary degree of freedom. They can be considered as composite fields of some fundamental fermions, which interact with each other through some unknown forces. In the effective Lagrangian description for light mesons, the Nambu-Jona-Lasinio (NJL) model \cite{Nambu:1961tp,Nambu:1961fr} is a time-honored example. Similar scenarios can happen at high energy scales. Recently, the authors of \cite{Inagaki:2015eza,Inagaki:2016vkf} engaged the gauged NJL model with inflationary machinery in slow-roll approximation. The predictions of such model are also consistent with the Planck 2015 data. Recent investigations also show that it is possible to construct models in which the inflaton emerges as a composite state of a four-dimensional strongly coupled theory \cite{Channuie:2011rq,Bezrukov:2011mv,Channuie:2012bv}.

The aim of this work is to present a unified description of inflation and dark matter in the context of the effective NJL model. In order to achieve our unified scenario, we also incorporate gravity in the NJL model in which the interaction between gravitons and the fundamental fermions induce a non-minimal coupling of the composite scalar bosons $\Phi$ to gravity. In analogy with the NJL models for light mesons, the bound states of the fundamental fermions can be classified into scalar channel, pseudoscalar channel and etc., and we will use scalar channel to describe the inflaton and the pseudoscalar channel to describe the dark matter, respectively. The composite scalar field is heavy and decoupled from the low energy degrees of freedom, e.g. the Standard Model particles, while the pseudoscalar field, being a Goldstone mode of the chiral symmetry of the fundamental fermions, is light (massless at the chiral limit) and connected to the low energy physics. In this sense both the pseudoscalar and the Higgs field are messengers between the inflation scale and the electroweak scale. 

This paper is organized as follows: In Sec.\ref{sec2}, we take a short recap of the NJL model with one flavor of some fundamental fermions and then incorporate the effective model to gravity. This inherently induces a nonminimal coupling of the composite scalar sectors to gravity. In Sec.\ref{sec3}, we demonstrate how an NJL effective potential emerges and propose a cosmological scenario that unifies cosmic inflation and dark matter to a single framework. In the same section, we show how a conformal transformation shapes it to an inflaton-type potential. We then compute the inflationary parameters and confront them with recent Planck 2015 data. We present in our model a simple connection of physics from the high scales to low scales via renormalization group equations of the physical parameters and also estimate for model of dark matter the thermally-averaged relic abundance for both the light and heavy CD masses. Discussions and conclusions are given in the last section.

\section{The Nambu--Jona-Lasinio model with gravity}
\label{sec2}

Various candidates of the fundamental fermions may emerge in different circumstances. Let us stress once again that the underlying description of models we are going to discuss are formulated via the fundamental fermions. Among them are listed as follows: In technicolor models, a new strongly interacting gauge theory (technicolor) and additional fundamental fermions (technifermions) are successfully incorporated. A bilinear condensate of technifermions in vacuum dynamically breaks the electroweak symmetry and provides the gauge boson masses (see Ref.\cite{Hill:2002ap} for a review). Moreover, the fundamental fermions appear in the supersymetric model of particle interactions (see Ref.\cite{Martin:1997ns} for a review). In addition, Majorana fermions are also compelling candidates for the fundamental ones and received much attention not only in particle physics \cite{Bilenky:2010zza}. Recently, it is proposed in Refs.\cite{Xiong:2016fum,Xiong:2016mxu} that new fundamental fermions (dark fermions) can emerge in the Standard Model via a spin-charge separation procedure. In the present paper we consider an Nambu-Jona-Lasinio (NJL) model with one flavor of some fundamental fermions $\psi$ that may possibly be one (all) of the candidates mentioned above:  
\begin{equation}
\mathcal{L} = \bar{\psi} i \gamma \cdot \partial \psi + G [ (\bar{\psi}\psi)^2 + ( \bar{\psi} i \gamma_5 \psi)^2], ~~~(\bar{\psi}\psi)^2 \equiv \left( \sum_{a=1}^{N_c} \bar{\psi}^a \psi^a\right)^2,
\end{equation}
where $N_c$ is the number of hyper-colors and $G$ is the NJL coupling which corresponds some new pairing force at the scale $\Lambda$,  the cut-off of the NJL theory. (Note that there are three-momentum cut-off scheme and four-momentum cut-off scheme, which will be discussed in the next section.) Therefore the NJL model has a parameter set consisting of $(G, \Lambda, N_c)$. 


Following the usual bosonization procedure \cite{Hatsuda, Klevansky, Ebert:1997fc},  we introduce two real scalar fields $\varphi$ and $S$ through
\begin{equation}
\Phi = \text{Re}\Phi + i \, \text{Im}\Phi \equiv \frac{1}{\sqrt{2}} \, (\varphi + i \,S),
\end{equation}
and they will be connected to the scalar channel and the pseudoscalar channel of the NJL model, respectively. The scalar field $\varphi$ will play the role of the inflaton and the scalar field $S$ will be considered as a candidate for dark matter. This is not unusual in the sense that one of the dark matter candidates, axion, is also a pseudoscalar.  Note that this does not mean that all dark matter come from the pseudoscalar channel.  In Ref.\cite{NMSM}, two Majorana spinors are also introduced. However,  in the present paper we will focus on these two scalar fields and leave the Majorana spinors for further investigation. With the help of the field $\Phi$ the  NJL Lagrangian can be expressed as 
\begin{eqnarray} \label{NJL}
\mathcal{L}_{\textrm{\tiny{NJL}}}  &=&  Z_\Phi \, \partial_{\mu} \Phi \, \partial^{\mu} \Phi^* - V_{\textrm{eff}}(\Phi, \Phi^*)  \cr
&&\quad+ \bar{\psi} \big[ i \gamma^{\mu}\partial_{\mu}  - ( \textrm{Re}\Phi +  i \gamma^5 \textrm{Im}\Phi ) + \cdots \big]\, \psi. 
\end{eqnarray}
where we only give explicitly the scalar channel and pseudo-scalar channel for an one-``flavor" case. As it will be shown later, $\textrm{Re}\Phi $ and $\textrm{Im}\Phi$ are related to the bilinear expressions of the constituent fermions $\bar{\psi} \psi$ and $ i \bar{\psi} \gamma^5 \psi$, respectively, and the effective potential $V_{\textrm{eff}}(\Phi, \Phi^*)$ can be calculated via the bosonization procedure which will be given in the next section. The kinetic term $\partial_{\mu} \Phi \, \partial^{\mu} \Phi^*$ emerges due to renormalization effect and $Z_{\Phi}$ is the wave-function renormalization constant. At high scales $\mu  \rightarrow \Lambda$, we require that 
\begin{equation}
Z_{\Phi} \rightarrow 0, ~~~~~\textrm{when}~~\mu \rightarrow \Lambda,
\end{equation}
hence $\Phi$ is not a dynamical field any more at high scales close to the NJL cutoff $\Lambda$ and  one can integrate it out through its equation of motion to reproduce the NJL four-fermion interaction, since
\begin{equation}
\Phi = \frac{1}{M_G^2} ~\bar{\psi}_R \, \psi_L = G ~ \bar{\psi}_R \, \psi_L, 
\end{equation}  
where we have defined a mass scale $M_G^2 = 1/G$ and chiral fermions $\psi_{R, L} = 1/2 (1 \pm \gamma_5) \psi$. Near the scale $M_G^2$ the Lagrangian in (\ref{NJL}) is equivalent to the NJL one.  When the scale decreases such that $\mu \ll M_G$, the field $\Phi$ becomes dynamical (See for example Ref. \cite{Hill91} for details using the renormalization group analysis). There will also be dynamical masses for the composite particles and constituent fermions. Roughly speaking, the mass of the \lq\lq sigma\rq\rq\ field $\varphi$ and the \lq\lq pion\rq\rq\ field $S$ are related via
\begin{equation}
m_\varphi^2 = 4 m + m_s^2\,,\label{massm}
\end{equation}
where $m$ is the dynamical mass of the constituent fermion, satisfying the gap equation \cite{Hatsuda, Klevansky}
\begin{equation}
m = - 2G \langle \bar{\psi} \psi \rangle + m_0  =   i 2 G N_c \text{Tr} \, S_{\psi} + m_0
\end{equation}
with $m_0$ being the ``current" mass of the constituent fermion and $S_{\psi}$ its propagator.  At the chiral limit ($m_0 \rightarrow 0$) the ``pion" modes are Goldstone one with vanishing mass so we have 
\begin{equation}
m_\varphi = 2 m
\end{equation}
In our case, if $m_\varphi$ is huge, say $\sim 10^{13}$ GeV, then the dynamical mass generated for the constituent fermions, whatever they are, would also be the same order of magnitude, while the Goldstone modes could gain small masses $m_s$, say $ m_s \sim 100$ GeV to $\sim$TeV, hence might be dark matter candidate as shown in \cite{NMSM}. Note that the cut-off $\Lambda$ of our NJL inflation model could be the GUT scale $\sim 10^{16}$ GeV. 

\begin{figure*}
\begin{center}
 \includegraphics[width=0.32\linewidth]{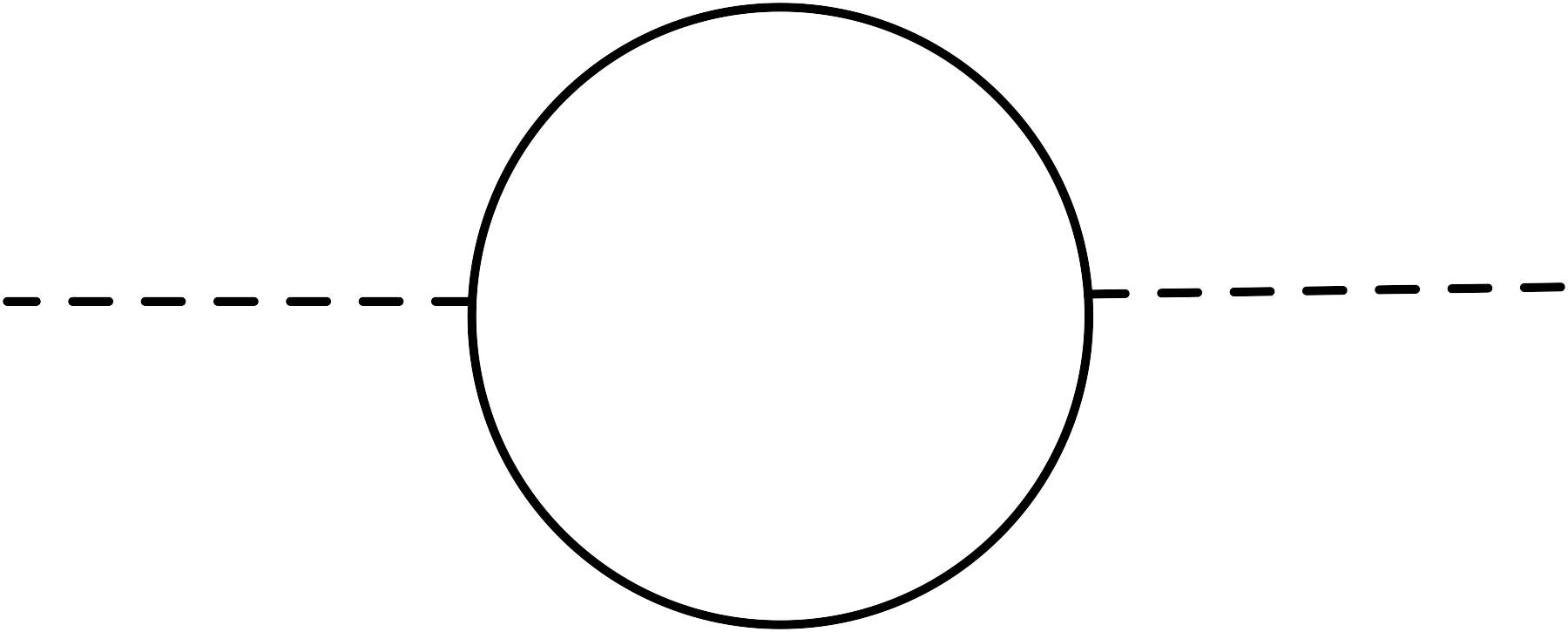}
  \includegraphics[width=0.23\linewidth]{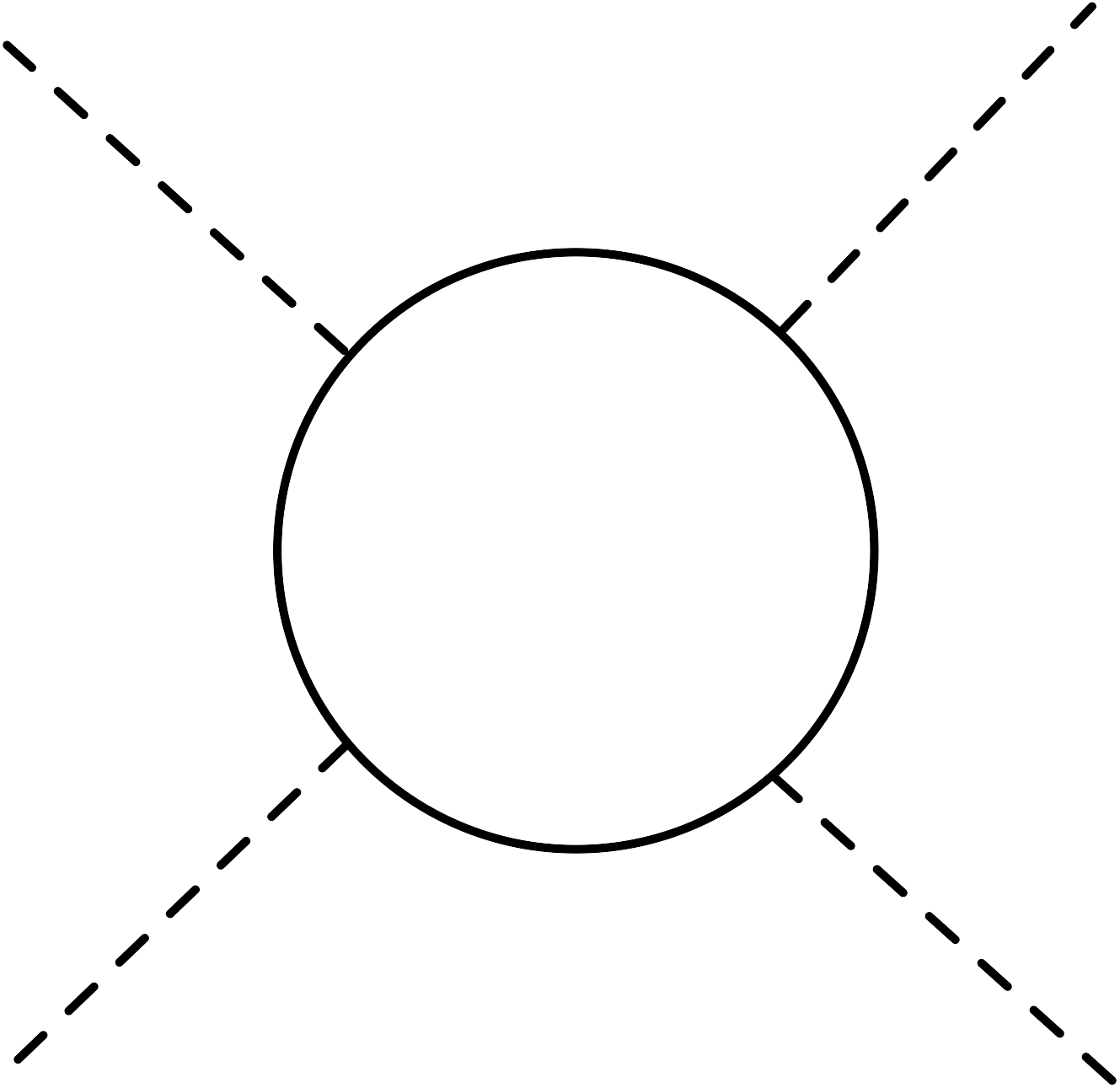}
  \includegraphics[width=0.28\linewidth]{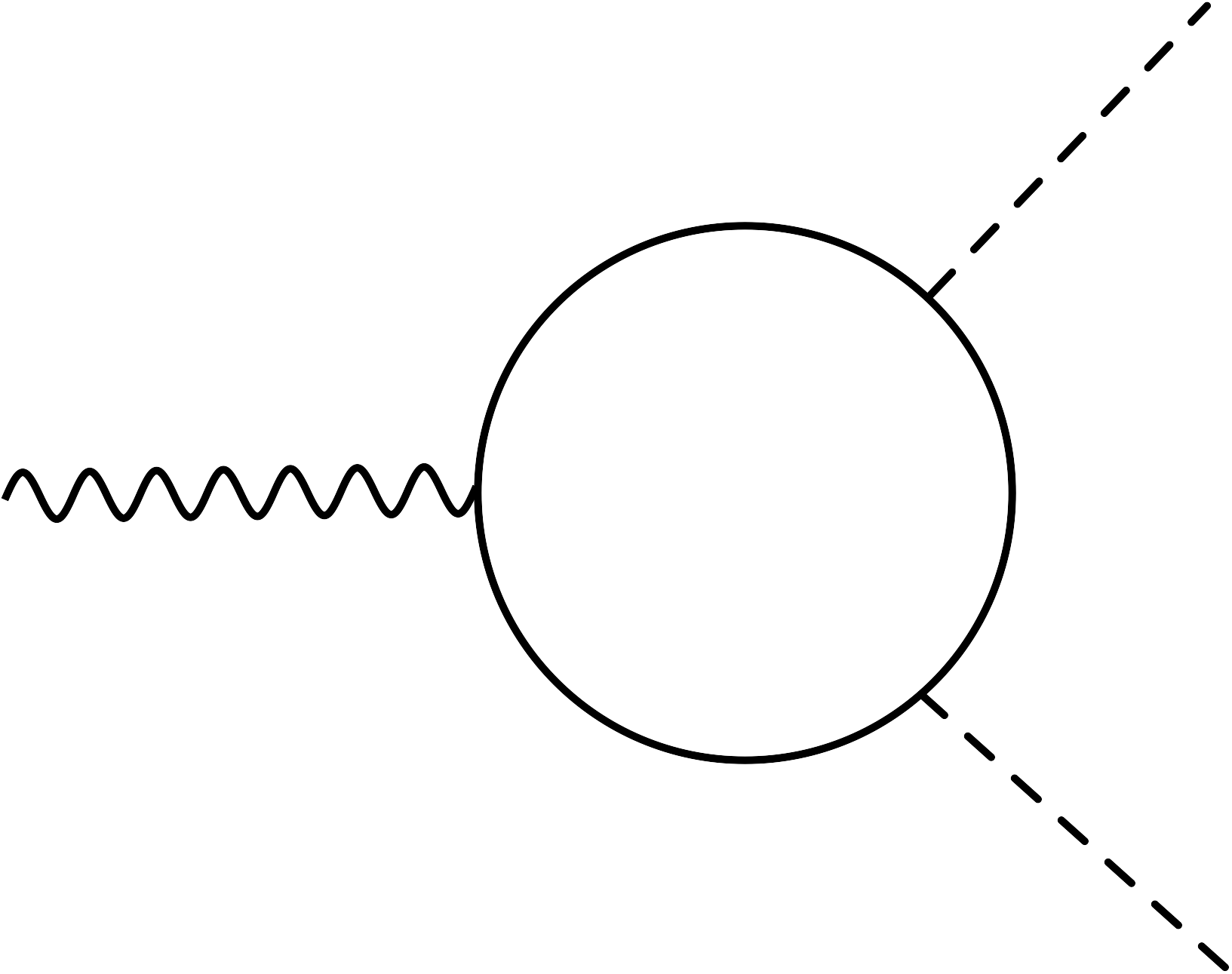}
 \caption{From left to right: the first two diagrams shows how the $Z_{\Phi}$ and $\lambda$ are induced by the fermion loops, respectively,  hence how the Eq.(\ref{NJL}) is obtained. The third diagram shows how the non-minimal coupling $\xi\Phi^{\dagger} \Phi R$ is induced by inserting the graviton. The solid lines indicate the fundamental fermion, the dashed lines are for the composite $\Phi$ field and the wavy line for the graviton. (The diagrams are reproduced from \cite{Hill91}).}
 \label{fig1}
\end{center}
\end{figure*}

Now we consider to incorporate gravity in the NJL model by placing it in some curved spacetime background with small gravitational fluctuations. The quanta of these fluctuations, gravitons, interact with the fundamental fermions and the absorption and emission of gravitons induce a
non-minimal coupling of the composite scalar bosons $\Phi$ to gravity, as shown by Hill and Salopek \cite{Hill91} (see Fig.\ref{fig1})
\begin{equation} \label{R-phi}
- \xi \, R \, \Phi^{\dagger} \Phi, 
\end{equation}
where $R$ is the Ricci scalar and $\xi$ is a coupling constant. They also found that $\xi = -1/6$ is an attractive renormalization group fixed point using the usual fermion bubble approximation (or random phase approximation). This is an interesting result since it has been well-known that  $\xi = -1/6$ coupling to gravity is conformal. A large non-minimal coupling is required in the Higgs-inflation scenario, $\xi \sim 10^4$. Nevertheless, as suggested in Ref.\cite{Hill91}, if $\xi$ is a running constant, its value at the NJL cutoff scale $\Lambda$, $\xi(\Lambda)$ might be large and then evolves toward $\xi = -1/6$ at low energies. One could also add other dimension 4 terms like $R^{2},\,R_{\mu\nu}R^{\mu\nu}$, etc., but they lead to terms with higher derivatives in the equations of motion and therefore, as also mentioned in Ref.\cite{Bezrukov:2008ut}, there exist additional degrees of freedom which should be dealt with in some special way. However, we do not consider such extensions in the present analysis\rq\rq. In our model we consider large-$\xi$ cases for the composite scalar and hence there is no conformal coupling region for the $\Phi$ field. We will also include the non-minimal coupling of the Higgs field to gravity (with coupling constant $\xi_h$). As it will be shown in the renormalization group analysis, even it is set to be zero at the electroweak scale, $\xi_h$ will become non-vanishing at high scales. 
The non-minimal coupling (\ref{R-phi}) leads to couplings of the scalars $\varphi$ and $S$ to the Ricci scalar 
\begin{equation}
- \frac{\xi}{2} \, R \, \varphi^2, ~~~~- \frac{\xi}{2} \, R \, S^2.
\end{equation}
Note that this is different from the Higgs-inflation models. Let us take the model in Ref.\cite{Clark09} as an example. Although both terms are considered in Ref.\cite{Clark09}, they come from different sources --- the inflaton comes from the Higgs doublet and the dark matter scalar is included by hand, hence their couplings to the Ricci scalar do not have to be the same.  In our model we use one parameter $\xi$ for describing the non-minimal coupling of both inflaton $\varphi$ and dark matter scalar $S$ to gravity, another parameter $\xi_h$ for the non-minimal coupling of the Higgs field to gravity and $\xi \gg \xi_h$. Nevertheless, similar to the Higgs-inflation models, the non-minimal coupling $ - \xi \, R \, \varphi^2 /2$ makes it possible for the NJL effective potential to successfully drive cosmic inflation. In the next section we will demonstrate how an NJL effective potential emerges and how a conformal transformation shapes it to an inflaton-type potential.

Note that in the Lagrangian (3) one can add a gauge field to make the derivative covariant with respective to some internal gauge symmetry. This has already been considered in Refs. \cite{Inagaki:2015eza,Inagaki:2016vkf} for composite inflation models. The gauged NJL models are more general in the sense that it can always reduce to the ungauged cases by switching off the gauge coupling constant. Nevertheless, it is not clear whether the gauge field should be included as a necessary ingredient. For instance the binding force for the fermions might be something else, not a gauge field; Also in the present work, for incorporating composite dark matter purpose, we do not intend to introduce extra gauge couplings. The gauged NJL model certainly has more degrees of freedom, but the question is whether it gives more model flexibility or dangerous physical processes, like unwanted coupling/decaying channels for dark matter (due to the possible coupling between the gauge field and dark matter). Even in the QCD case, phenomenologically a gauged NJL model does not seem to add much to the ungauged one \cite{Hatsuda, Klevansky} -- the electromagnetic fields generated in heavy-ion collisions are too weak to affect the chiral symmetry breaking \cite{Klevansky}; For color electric and magnetic fields,  they might be relevant in addressing confinement and topological charge problems \cite{Xiong:2013uuf,Xiong:2014yba}. However, whether the composite inflation (and dark matter) model should have confinement or not is unknown. Therefore in the present paper we will be satisfied if the NJL model with the simplest configuration is consistent with experimental observations, and refer to Refs. \cite{Inagaki:2015eza,Inagaki:2016vkf} (on inflation) for the gauged NJL studies and leave them to future investigations (on inflation and dark matter).

\section{A composite model for inflation and dark matter}
\label{sec3}

The effective Lagrangian of NJL models can be calculated via the path integral approach (for a review, see e.g. \cite{Ebert:1997fc}). The generating functional $\mathcal{Z}$ of the NJL model can be used to identify an effective Lagrangian  through
\begin{equation}
\mathcal{Z} = \int [\mathcal{D} \psi \mathcal{D} \bar{\psi}] ~ e^{ i \int d^4x \mathcal{L}_{\textrm{\tiny{NJL}}}}  \longrightarrow   \int [\mathcal{D}\varphi  \mathcal{D} S \cdots ]  ~ e^{ i \int d^4x \mathcal{L}_{\textrm{\tiny{eff}}}}
\end{equation}
where the ellipsis stands for the other collective ``meson" fields. (see a review on the NJL in curved spacetime \cite{Inagaki:1997kz}.) This bosonization procedure starts with replacing the NJL four-fermion interaction by a Yukawa-type coupling at the tree level, with the help of an auxiliary complex scalar $\Phi$
\begin{equation}
\Phi = \frac{1}{M_G^2} ~\bar{\psi}_R \, \psi_L = G ~ \bar{\psi}_R \, \psi_L\,.
\end{equation}  
The action now becomes
\begin{eqnarray} \label{NJLcurved}
{\cal S} &=& \int d^4x \sqrt{-g} \, \bigg[ - \frac{1}{2} M^{2}_{\textrm{\tiny{PL}}} R  - M^2_{G} \Phi^{\dagger} \Phi   \cr
 &&\quad\quad\quad\quad\quad\quad + \bar{\psi} \big[ i \gamma^{\mu}(\nabla_{\mu}  - ( \textrm{Re}\Phi +  i \gamma^5 \textrm{Im}\Phi )  \big] \, \psi \,\, + \cdots \bigg] 
\end{eqnarray}
The kinetic term of the $\Phi$ field will be introduced later. At this stage (or scales close to the NJL cutoff $\Lambda$), the field $\Phi$ is an auxiliary field instead of a dynamical one. One can integrate it out through its equation of motion and reproduce the NJL four-fermion interaction in the curved space. Near the scale $M_G^2$ the Lagrangian in (\ref{NJLcurved}) is equivalent to the NJL one. When gravity is involved, the fermion loop will induce a coupling between the curvature term $R$ and the field $\Phi$ (see the third diagram in Fig.1),
\begin{equation}
- \xi \, R\,  \Phi^{\dagger} \Phi ,
\end{equation}
where the coefficient $\xi$ was found to be the conformal one ($\xi = -1/6$) in ref. \cite{Hill91} but we will consider more general cases and treat $\xi$ as an arbitrary parameter. Now we have the relevant ingredients of the composite field $\Phi$ for our investigation: 
\begin{eqnarray} \label{NJLcurvedxi}
{\cal S} = \int d^4x \sqrt{-g} \, \bigg[ - \frac{1}{2} M^{2}_{\textrm{\tiny{PL}}} R - \xi \, R\,  (\Phi^{\dagger} \Phi - \frac{v^2}{2}) + g^{\mu\nu}\partial_{\mu}\Phi \partial_{\nu}\Phi^{\dagger}- V_{\textrm{\tiny{eff}}} ( \Phi^{\dagger},  \Phi) \bigg] \
\end{eqnarray}
where we have included an effective potential for the scalar field $\Phi$, which is simply a Higgs-type one
\begin{equation}
V_{\textrm{\tiny{eff}}} ( \Phi^{\dagger},  \Phi) = - \mu_0^2 \Phi^{\dagger} \Phi + \frac{1}{2} \lambda \, (\Phi^{\dagger} \Phi)^2,
\end{equation}
and the mass term $ M^2_{G} \Phi^{\dagger} \Phi $ has been absorbed into the effective potential. The fact that the effective Lagrangian (\ref{NJLcurvedxi}) can be derived from an NJL model demonstrates a strong resemblance to the relation between the Ginzburg-Landau phenomenological model  and the BCS theory of superconductivity \cite{Klevansky} --- the Ginzburg-Landau theory can be derived as an effective model from the more fundamental BCS theory. Here we follow a more explicit derivation of the Higgs-type potential from Ref.\cite{Ebert:1997fc}, based on the bosonization procedure to obtain the composite fields $\varphi$ and $S$. The effective Lagrangian of $\Phi$ was obtained in terms of real scalar fields $\varphi (x)$ and $S(x)$ ( writing $\Phi (x) = \varphi (x) + i S (x)$ )
\begin{eqnarray}
\mathcal{L}_{\textrm{\tiny{eff}}} &=&  \frac{1}{2} \partial_\mu \varphi  \partial^\mu \varphi + \frac{1}{2} \partial_\mu S \partial^\mu S - \frac{1}{2} m^2_\varphi \varphi^2 - \frac{1}{2} m^2_s S^2 \cr
&& - g_{\varphi s s} \varphi (\varphi^2 + S^2) - g_{4s} (\varphi^2 + S^2)^2
\end{eqnarray}
where the masses and coupling constants are found to be \cite{Ebert:1997fc}
\begin{eqnarray} \label{NJLset2}
m^2_s &=& m_0 \frac{g^2_{s\psi\psi}}{m \,G}, ~~~m^2_\varphi = m^2_s + 4 m^2, \cr
 g^2_{s\psi\psi} &=& \frac{1}{2 \sqrt{N_c I_2}}, ~~~g_{\varphi s s} = \frac{m}{\sqrt{N_c I_2}}, ~~~g_{4s} = \frac{1}{8 N_c I_2}, 
\end{eqnarray}
where $g_{4s}$ will be identified with $\lambda$, i.e. $g_{4s} = \lambda/2$ and the integral $I_2 (\Lambda, m) $ is given by
\begin{equation}
 I_{2} = -i\int^{\Lambda}_0 \frac{d^{4}k}{(2\pi)^{4}}\frac{1}{\left(k^{2} - m^{2}\right)^{2}}.
\end{equation}
The trilinear terms $g_{\varphi s s} \varphi (\varphi^2 + S^2)$ complicate the story. For simplicity we introduce a discrete symmetry $Z_2$ for the inflaton field $\varphi$:
\begin{equation}
\varphi \longrightarrow - \varphi
\end{equation}
under which the Lagrangian is asked to be invariant. This allows us to remove the tri-linear couplings. Similarly when the couplings to the Higgs particle are concerned, one can also impose a $Z_2$ symmetry on the $S$ field to forbid the decay channel $S \rightarrow HH$. 

The action (\ref{NJLcurvedxi}) can basically be recast in terms of the component fields $\varphi$ and $S$:
\begin{eqnarray} \label{NJLcurvedxi1}
{\cal S}_{\textrm{\tiny{NJL}}} = \int d^4x \sqrt{-g} \, \bigg[ &-& \frac{1}{2} M^{2}_{\textrm{\tiny{PL}}} R + \frac{1}{2}g^{\mu\nu}\partial_{\mu}\varphi\partial_{\nu}\varphi +\frac{1}{2}g^{\mu\nu}\partial_{\mu}S\partial_{\nu}S - \frac{\xi \, R}{2}\,  \left(\varphi^{2}+S^{2} - \frac{v^2}{2}\right)  \cr &-& \frac{1}{2} m^{2}_{\varphi} \varphi^{2}-   \frac{1}{2} m^{2}_{S} S^{2} -\frac{1}{2}\lambda\left(\varphi^{2}+S^{2}\right)^{2}  \bigg]. \
\end{eqnarray}
To connect with the Standard Model at the electroweak scale, we need to take into account the couplings between the $\Phi$ field and the Standard Model particles such as the Higgs fields. There could be some unknown couplings between the fundamental fermion fields $\psi$ and the Higgs fields. For instance in Ref. \cite{Xiong:2016fum,Xiong:2016mxu} it has been suggested that the leptons can be considered as bound states of some fundamental fermions and Higgs fields (Bose-Fermi mixture), while the pairing of the fundamental fermions leads to the dark matter.  Therefore it is reasonable to include a coupling
\begin{equation}
-\kappa \Phi^{\dagger} \Phi H^{\dagger} H = -\frac{1}{4} \kappa  (\varphi^2 + S^2) h^2
\end{equation}
to describe how the composite scalar $\Phi$ interacts with the Standard Model Higgs $H = \frac{1}{\sqrt{2}} (0, ~ h+v_0)^T$ (unitary gauge). Furthermore one can write down the Higgs potential and its possible couplings with the Ricci scalar, 
\begin{equation}
\lambda_h \left( H^{\dagger} H - \frac{v^2_0}{2} \right)^2 = \frac{1}{2} m^2_h h^2 + \sqrt{\frac{\lambda_h}{2}} m_h h^3 + \frac{1}{4}\lambda_h h^4  , ~~~~~ - \xi_h  H^{\dagger} H R = - \frac{1}{2} \xi_h h^2 R , 
\end{equation}
where inflation models based on the non-minimal couplings of the Higgs field and a real or complex scalar field to the Ricci scalar, respectively, have been studied in Refs. \cite{Clark09, Lerner:2009xg}. In these models the role of dark matter is played by the real or complex scalar field (gauge singlet). Similar to Ref.\cite{Lerner:2009xg}, we will focus on inflation along the ``$\varphi$-direction" in which the non-minimal coupling between the composite fields and gravity dominates over the non-minimal coupling for the Higgs field ($\xi \gg \xi_h)$.

\subsection{Composite inflaton (CI) from the NJL}
Notice that both $\varphi$ and $S$ has the same (non-minimal) coupling, $\xi$, to gravity. At very high energy scale, the mean field of $S$ is very small compared with that of the field $\varphi$. Another word of saying, the field $S$ is supposed to be a massless mode at very high energy scale. Therefore in order to examine model of inflation, we can now suppress the contribution from $S$ and consider the $\varphi$ dynamics only. Therefore the action describing model of inflation in the Jordan (J) frame reads
\begin{eqnarray} \label{NJLcurvedxi2}
{\cal S}^{\rm{J}}_{\textrm{\tiny{CI}}} &=& \int d^4x \sqrt{-g} \, \bigg[ - \frac{1}{2} M^{2}_{\textrm{\tiny{PL}}} R + \frac{1}{2}g^{\mu\nu}\partial_{\mu}\varphi\partial_{\nu}\varphi - \frac{\xi \, R}{2}\,  \left(\varphi^{2} - \frac{v^2}{2}\right)  - V_{\textrm{\tiny{eff}}}(\varphi) \bigg]\,, \cr
V_{\textrm{\tiny{eff}}}(\varphi) &=& - \frac{1}{2} m^{2}_{\varphi}\varphi^{2} + \frac{1}{2}\lambda\varphi^{4}.
\end{eqnarray}
It was noticed so far in the framework of Higgs-inflation investigated in Ref.\cite{Bezrukov:2007ep} that the non-minimal coupling ($\xi$) of the Higgs doublet field $(H)$ to gravity, i.g. $\sim\xi H^{\dagger}HR$,  has the salient feature. The reason resides from the fact that a nonzero value of $\xi$ is needed since for $\xi=0$ an unacceptably large amplitude of primordial inhomogeneities is generated for a realistic quartic Higgs self-interaction term \cite{Bezrukov:2008ut}. Specifically, it was found in \cite{Bezrukov:2007ep} that with $\xi$ of the order $10^{4}$ the model leads to successful inflation and produces the spectrum of primordial fluctuations in good agreement with the observational data.

Due to the presence of the non-minimal coupling term phenomenologically introduced, it is more convenient to diagonalize into another form by applying a conformal transformation. This allows us to rewrite the action as minimally coupled but with a new canonically normalized field. Hence the conformal transformation can be basically implemented by making use of the following replacement:
\begin{equation}
\tilde{g}_{\mu\nu} = \Omega^{2}\,g_{\mu\nu} = \left(1 + \frac{\xi\,(\varphi^{2} - v^2/2)}{M^{2}_{\textrm{\tiny{PL}}}}\right)g_{\mu\nu}\,.
\end{equation}
Thus the action in (\ref{NJLcurvedxi2}) becomes the Einstein-frame (E) form:
\begin{eqnarray} \label{NJLEif}
{\cal S}^{\rm E}_{\textrm{\tiny{CI}}} = \int d^4x \sqrt{-g} \, \bigg[ -\frac{1}{2} M^{2}_{\textrm{\tiny{PL}}} R + \frac{1}{2}\Omega^{-4}\left(\Omega^{2} + \frac{6\xi\varphi^{2}}{M^2_{\textrm{\tiny{PL}}}}\right)g^{\mu\nu}\partial_{\mu}\varphi\, \partial_{\nu}\varphi - U_{\rm {\tiny{eff}}}(\varphi)\bigg] \,,
\end{eqnarray}
where
\begin{equation} \label{UEi}
\Omega^{2} = \left(1 + \frac{\xi\,(\varphi^{2} - v^2/2)}{M^{2}_{\textrm{\tiny{PL}}}}\right)\quad{\rm and}\quad U_{\rm {\tiny{eff}}}(\varphi)\equiv\Omega^{-4}V_{\textrm{\tiny{eff}}}(\varphi)\,.
\end{equation}
However, the transformation leads to a non-canonical kinetic term for the scalar field. It is convenient to put in a canonical form by introducing a new canonically normalized scalar field $\chi$ satisfying the relation
\begin{eqnarray} \label{canonical}
\frac{1}{2}g^{\mu\nu}\partial_{\mu}\chi(\varphi)\partial_{\nu}\chi(\varphi) = \frac{1}{2}\left(\frac{d\chi}{d\varphi}\right)^{2}g^{\mu\nu}\partial_{\mu}\varphi\partial_{\nu}\varphi\,,
\end{eqnarray}
where
\begin{eqnarray} \label{fieldchi}
\chi^{\prime} = \left(\frac{d\chi}{d\varphi}\right) = \sqrt{\Omega^{-4}\left(\Omega^{2} + \frac{6\xi\varphi^{2}}{M^{2}_{\textrm{\tiny{PL}}}}\right)}\,.
\end{eqnarray}
It is noticed that, for small field value, i.e. $\xi\varphi^{2}\ll M^{2}_{\textrm{\tiny{PL}}}$, the potential for the field $\chi$ is the same as that of the original field, $\varphi$. However, it is not the case for large value of the field, i.e. $\xi\varphi^{2}\gg M^{2}_{\textrm{\tiny{PL}}}$. In the later case, we find the solution of $\varphi$ written in terms of the field $\chi$ as
\begin{eqnarray} \label{soluphi}
\varphi \simeq \frac{M_{\textrm{\tiny{PL}}}}{\sqrt{\xi}}\exp\left(\frac{\chi}{\sqrt{6}M_{\textrm{\tiny{PL}}}}\right)\,.
\end{eqnarray}
The effective potential $U_{\rm {\tiny{eff}}}(\chi)$ has the form
\begin{eqnarray} \label{solpot}
U_{\rm {\tiny{eff}}}(\chi) \simeq \frac{\lambda M^{4}_{\textrm{\tiny{PL}}}}{2\xi^{2}}\bigg[1+ \exp\left(-\frac{2\chi}{\sqrt{6}M_{\textrm{\tiny{PL}}}}\right)\bigg]^{-2}\,,
\end{eqnarray}
where we have also imposed the limit in which the field is far away from the minimum of its potential such that $\xi v^{2}\ll M^{2}_{\textrm{\tiny{PL}}}$. In the limit of $\varphi^{2}\gg M^{2}_{\textrm{\tiny{PL}}}/\xi\gg v^{2}$, the slow-roll parameters \cite{Lyth:1998xn} in the Einstein frame can be expressed as functions of the field $\varphi(\chi)$:
\begin{eqnarray}
\epsilon   & := & \frac{M^{2}_{\textrm{\tiny{PL}}}}{2}\left(\frac{dU_{\rm {\tiny{eff}}}/d\chi}{U_{\rm {\tiny{eff}}}}\right)^{2} = \frac{M^{2}_{\textrm{\tiny{PL}}}}{2}\left(\frac{U_{\rm {\tiny{eff}}}'}{U_{\rm {\tiny{eff}}}}\frac{1}{\chi'}\right)^{2} \simeq \frac{4M^{4}_{\textrm{\tiny{PL}}}}{3\xi^{2}\varphi^{4}}, \cr
\eta & := & M^{2}_{\textrm{\tiny{PL}}}\frac{d^{2}U_{\rm {\tiny{eff}}}/d\chi^{2}}{U_{\rm {\tiny{eff}}}} = M^{2}_{\textrm{\tiny{PL}}}\frac{U_{\rm {\tiny{eff}}}''\chi' - U_{\rm {\tiny{eff}}}\chi''}{U_{\rm {\tiny{eff}}}\chi'^{3}} \simeq -\frac{4M^{2}_{\textrm{\tiny{PL}}}}{3\xi\varphi^{2}}, \cr
\zeta & := & M^{4}_{\textrm{\tiny{PL}}}\frac{\left(d^{3}U_{\rm {\tiny{eff}}}/d\chi^{3}\right)dU_{\rm {\tiny{eff}}}/d\chi}{U_{\rm {\tiny{eff}}}^{2}} \simeq \frac{16M^{4}_{\textrm{\tiny{PL}}}}{9\xi^{2}\varphi^{4}}\,,
\end{eqnarray}
where \lq\lq\,\,$'$\,\,\rq\rq\,denotes derivative with respect to $\varphi$. Notice that the results we obtained here are approximately the same for those of inflationary model driven by the SM Higgs boson \cite{Bezrukov:2007ep}. Slow-roll inflation terminates when $\epsilon = 1$, so the field value at the end of inflation reads $\varphi_{\rm {\tiny{end}}} \simeq (4/3)^{1/4}M_{\textrm{\tiny{PL}}}/\sqrt{\xi}$. The number of e-foldings for the change of the field $\varphi$ from $\varphi_{N}$ to $\varphi_{\rm {\tiny{end}}}$ is given by
\begin{eqnarray} \label{efold}
N = \frac{1}{M^{2}_{\textrm{\tiny{PL}}}} \int^{\chi_{N}}_{\chi_{\rm {\tiny{end}}}} \frac{U_{\rm {\tiny{eff}}}}{dU_{\rm {\tiny{eff}}}/d\chi}d\chi = \frac{1}{M^{2}_{\textrm{\tiny{PL}}}} \int^{\varphi_{N}}_{\varphi_{\rm {\tiny{end}}}} \frac{U_{\rm {\tiny{eff}}}}{dU_{\rm {\tiny{eff}}}/d\varphi}\left(\frac{d\chi}{d\varphi}\right)^{2}d\varphi \simeq \frac{6\xi}{8M^{2}_{\textrm{\tiny{PL}}}}\left(\varphi^{2}_{N} - \varphi^{2}_{\rm {\tiny{end}}}\right)\,,
\end{eqnarray}
where $\varphi_{N}$ represents the field value corresponding to the horizon crossing of the observed CMB modes. After substituting $\varphi_{\rm {\tiny{end}}}$ into the above relation, we obtain $\varphi_{N} \simeq 9 M_{\textrm{\tiny{PL}}}/\sqrt{\xi}$ for $N = 60$. To generate the proper amplitude of the density perturbations, the potential must satisfy the COBE renormalization $U_{\rm {\tiny{eff}}}/\epsilon \simeq (0.0276\,M_{\textrm{\tiny{PL}}})^{4}$ \cite{Bezrukov:2008ut}. Inserting (\ref{UEi}) and (\ref{efold}) into the COBE normalization, we find the required value for $\xi$
\begin{eqnarray} \label{xi}
\xi \simeq \sqrt{\frac{2\lambda}{3}}\frac{N}{(0.0276)^{2}}\,.
\end{eqnarray} 
\begin{figure*}
\begin{center}
 \includegraphics[width=0.47\linewidth]{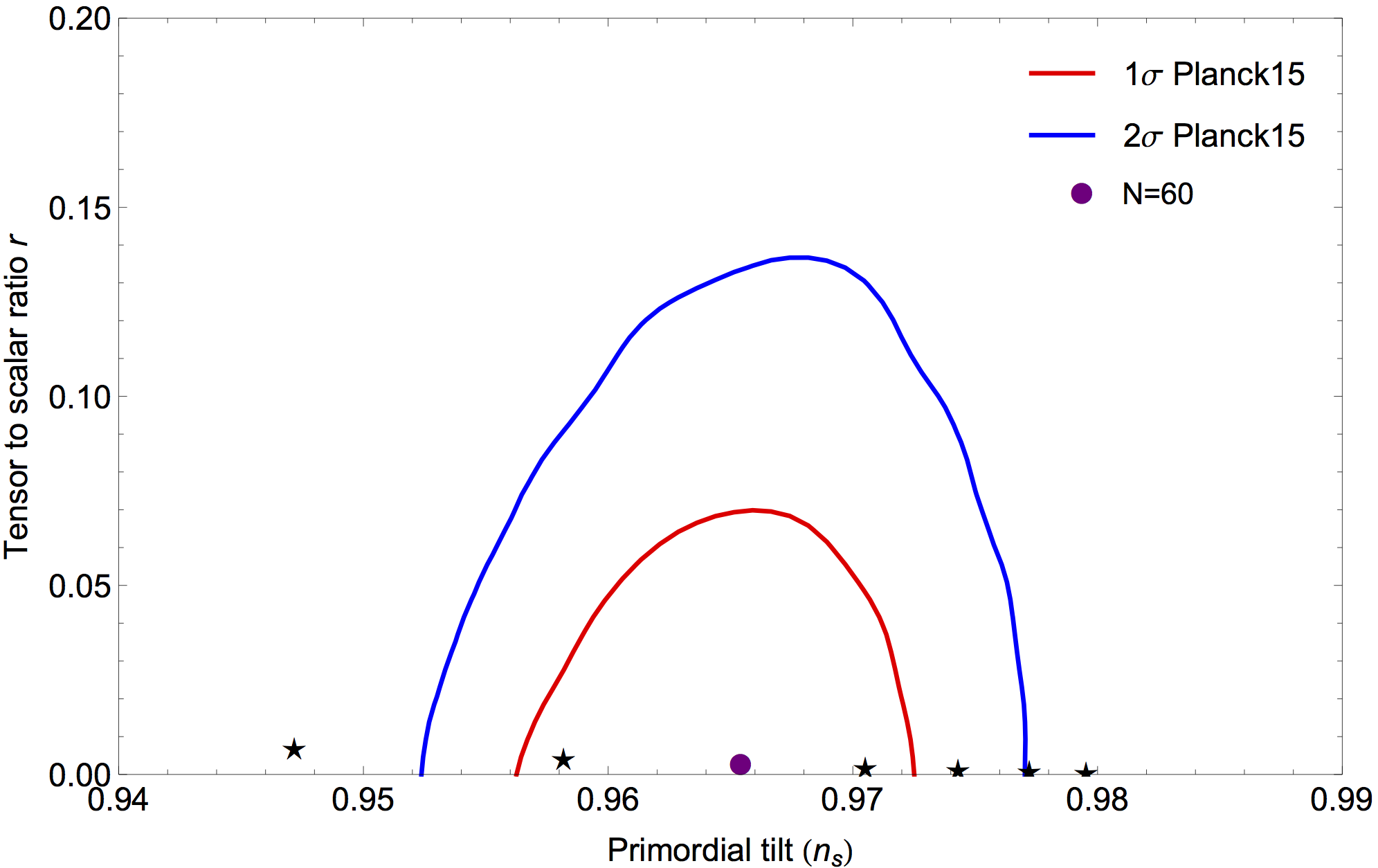}
  \includegraphics[width=0.49\linewidth]{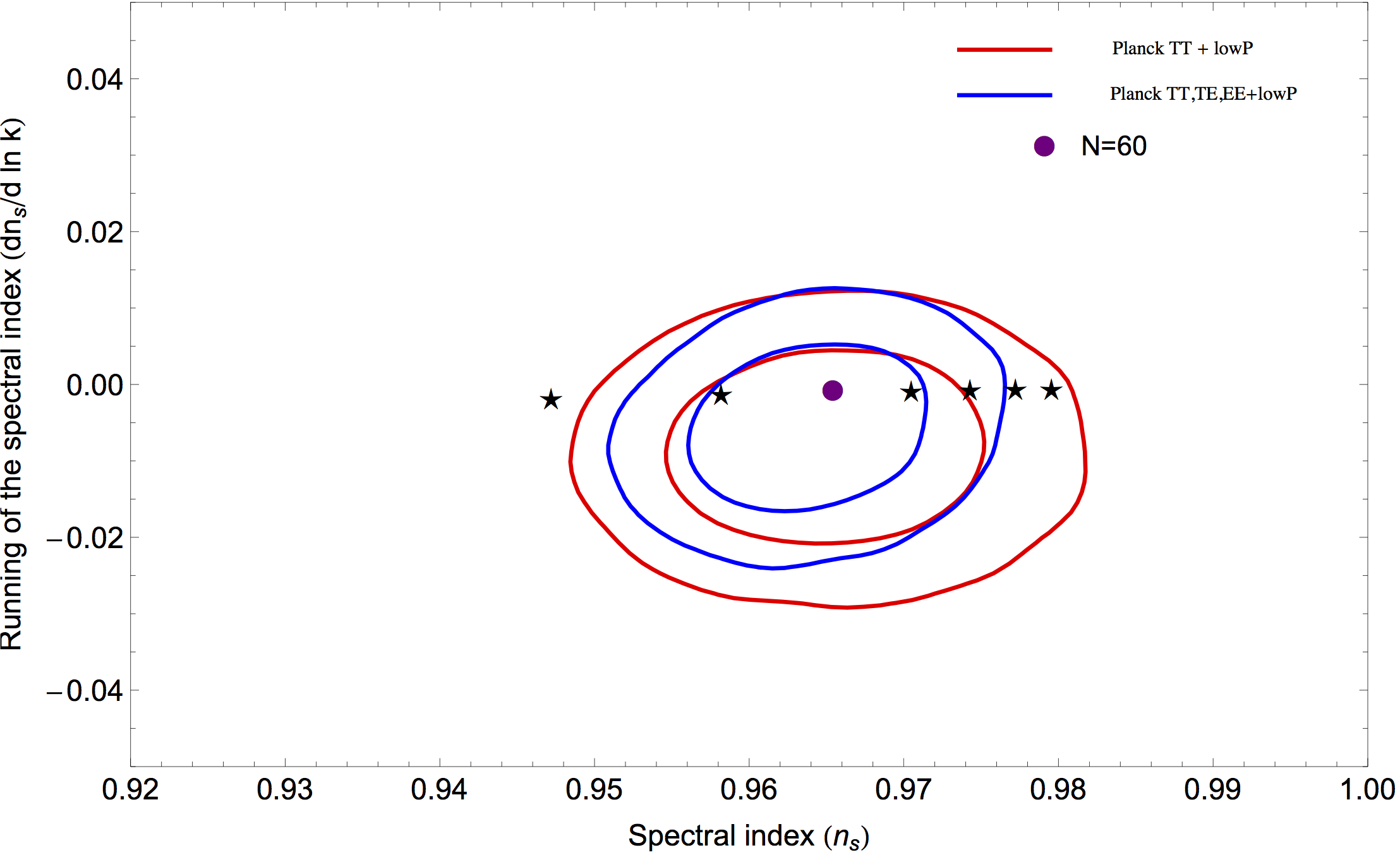}
 \caption{Left panel: We compare the theoretical predictions in the $(r-n_{s})$ plane for different values of e-folds $N$ with Planck$'15$ results for TT, TE, EE, +lowP and assuming $\Lambda$CDM + r \cite{Ade:2015lrj}; Right panel: Marginalized joint $68\%$ and $95\%$\,C.L. for ($n_{s},\,dn_{s}/d ln k$) using Planck TT+lowP and Planck TT,TE,EE+lowP. For comparison, Fig.\,\ref{fig11} shows the predictions for this model with (from left to right) $N=[40,100]$ \cite{Ade:2015lrj}.}
 \label{fig11}
\end{center}
\end{figure*}
\begin{figure*}
\begin{center}
 \includegraphics[width=0.57\linewidth]{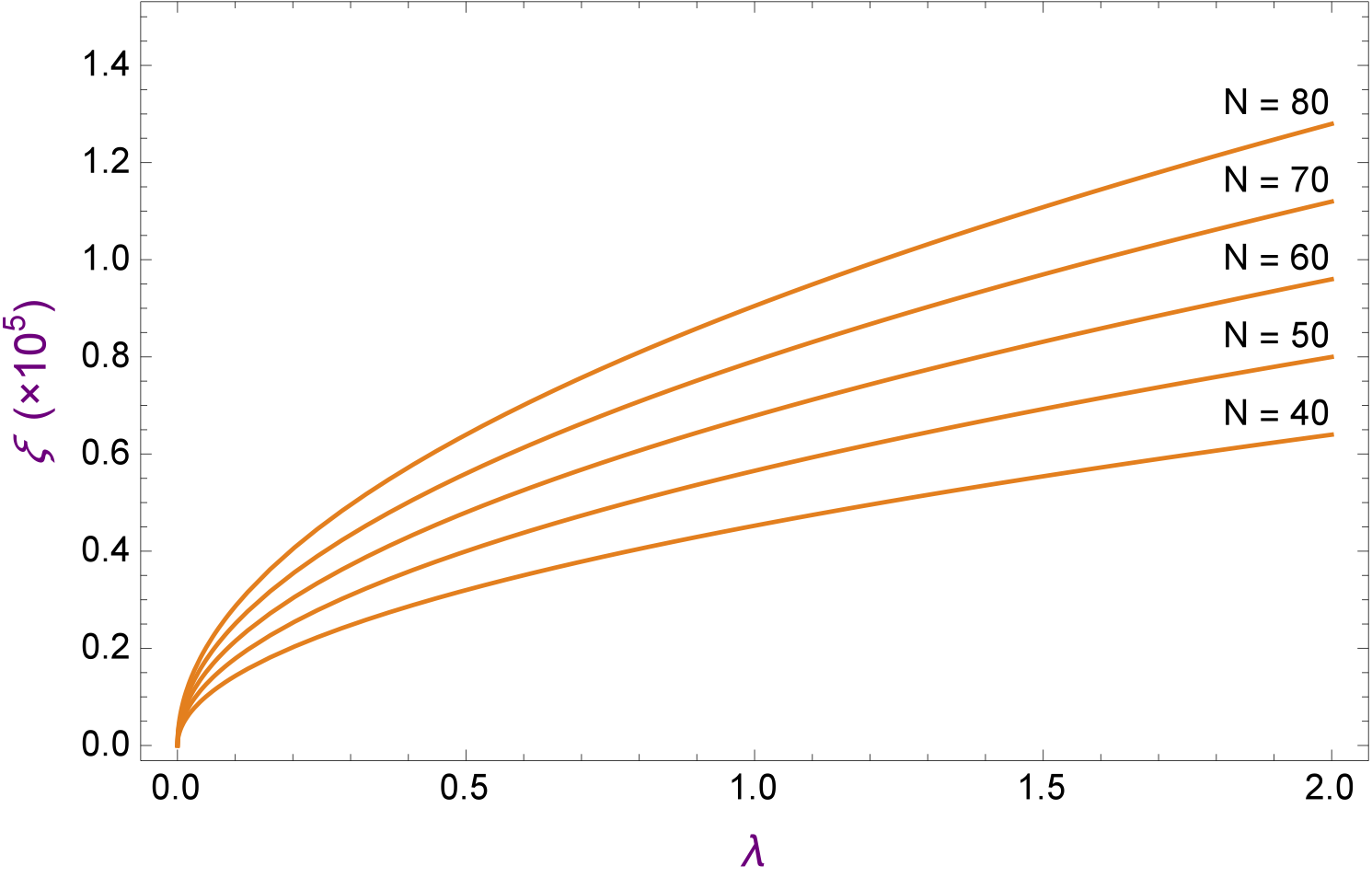}
 \caption{This figure shows the non-minimal coupling $\xi$ dependence on the self-coupling $\lambda$ for different values of the e-foldings $N$.}
 \label{fig4}
\end{center}
\end{figure*}
To the lowest order in $1/\xi$, the amplitude of the power spectrum for the curvature perturbations $A_{s}$ reads 
\begin{eqnarray} \label{Aparas}
A_{s}  :=  \frac{U_{\rm {\tiny{eff}}}}{24\pi^{2}M^4_{\textrm{\tiny{PL}}}\epsilon} \simeq  \frac{\lambda {N}^{2}}{36\pi^{2}\xi^{2}} \simeq 2.2\times 10^{-9},
\end{eqnarray}
and the spectral index of curvature perturbation $n_{s}$ and its running $n'_{s}$, and the tensor-to-scalar ratio $r$ are given in terms of the e-foldings $N$:
\begin{eqnarray} \label{paras}
n_{s} & := & 1- 6\epsilon +2\eta \simeq 1-\frac{2}{N} - \frac{9}{2{N^{2}}}, \cr
n'_{s} & := & dn_{s}/d\ln k \simeq -\frac{2}{N^{2}} - \frac{12}{N^{3}} - \frac{27}{2N^{4}}, \cr
r & := & 16\epsilon \simeq \frac{12}{N^{2}} .
\end{eqnarray}
\begin{figure*}
\begin{center}
 \includegraphics[width=0.47\linewidth]{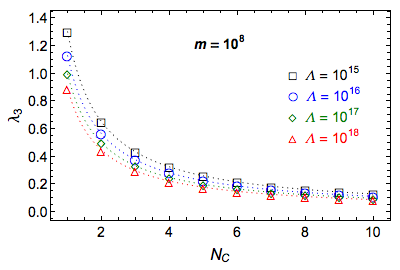}
  \includegraphics[width=0.47\linewidth]{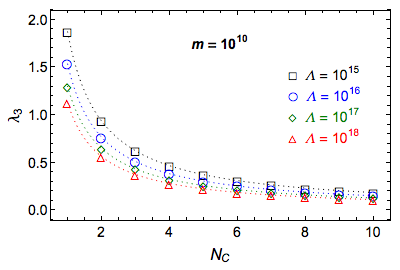}
   \includegraphics[width=0.47\linewidth]{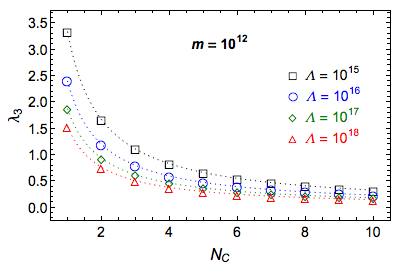}
    \includegraphics[width=0.47\linewidth]{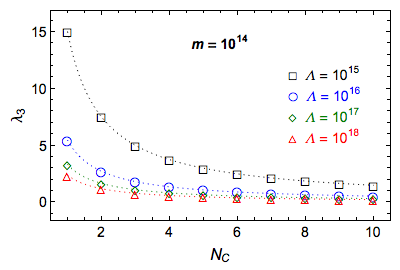}
 \caption{The self-coupling $\lambda$ dependence on $N_{c}$ with the 3-momentum cutoff scheme $(\lambda := \lambda_{3})$ for different values of the NJL parameters $m$ and $\Lambda$, based on the 3MCS case of Eq.(\ref{scheme}).}
 \label{fig2}
\end{center}
\end{figure*}
Here we compared our results with the recent Planck measurement by placing the predictions in the ($r-n_{s}$) plane with different values of e-folds, $N$, illustrated in Fig.(\ref{fig11}). We find that in order to lie within $1\sigma$ C.L. of Planck 2015 contours the number of e-folds should satisfy $48 \lesssim N\lesssim 75$. For example, we obtain from Eq.(\ref{xi}) and (\ref{paras}) that $\xi\sim 64,000\sqrt{\lambda},\,n_{s}\simeq 0.966$ and $r\simeq 0.0033$ for $N=60$ e-folds. From Fig.(\ref{fig11}), we find for this model that the running of the scalar spectral index does not significantly change as a function of $n_{s}$. Considering Eq.({\ref{xi}}) allows us to demonstrate the non-minimal coupling $\xi$ dependence on the self-coupling $\lambda$ for different values of the e-foldings $N$ illustrated in Fig.\ref{fig4}.

It would be a great deal of interest in relating the self-coupling $\lambda$ with the parameters from the NJL model, e.g. the cutt-off $\Lambda$ and the number of color $N_{c}$. To begin with, we start by using the results given in Ref.\,\cite{Ebert:1997fc} and we find for this work
\begin{eqnarray} \label{g}
g_{4s}=\frac{\lambda}{2} = \frac{1}{8 N_{c} I_{2}}\,,
\end{eqnarray} 
where 
\begin{eqnarray}
I_{2} = -i\int^{\Lambda} \frac{d^{4}k}{(2\pi)^{4}}\frac{1}{\left(k^{2} - m^{2}\right)^{2}}\,,\,\,{\rm with}\,\,m^{2}_{\varphi} = m^{2}_{s}+4m^{2}. \label{I2}
\end{eqnarray} 
After performing the above integration, then cosmological parameters can be rewritten in terms of the NJL information. Notice that, however, there are two different cutoff schemes \cite{Hatsuda, Klevansky} : the 3-momentum cutoff scheme (3MCS) and the 4-momentum cutoff scheme (4MCS). Hence, from Eq.(\ref{I2}) we find
\begin{equation} \label{scheme}
 I_{2} = \begin{cases}
        \frac{1}{16 \pi^2} \left[ - \frac{\Lambda^2}{\Lambda^2+m^2} +\ln\left(1 + \frac{\Lambda^2}{m^2}  \right) \right], & \text{for the 4MCS,}\\
        &\\
        \frac{1}{8 \pi^2} \left[ - \frac{\Lambda_3}{\sqrt{\Lambda^2_3+m^2}} +\ln\left(\frac{\Lambda_3+\sqrt{\Lambda^2_3+m^2}}{m}  \right) \right]. & \text{for the 3MCS.}
        \end{cases}
\end{equation}
If $\Lambda \gg m $, and $\Lambda_3 \gg m$, then in both schemes we obtain the same result
\begin{equation}
I_2 \approx \frac{1}{8 \pi^2} \ln\left(\frac{\Lambda}{m} \right)\,,
\end{equation}
and hence the relation between $\lambda$ and the NJL parameters
\begin{equation} \label{3or4}
\lambda \approx \frac{2 \pi^2}{N_c \ln \frac{\Lambda}{m}}
\end{equation}
where we have used a relation from Ref.\,\cite{Hatsuda}  $\Lambda^2_3 = (\Lambda/2)^2 - m^2 $.
\begin{figure*}
\begin{center}
 \includegraphics[width=0.47\linewidth]{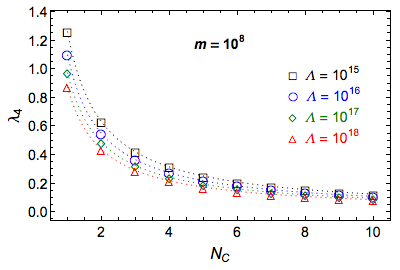}
  \includegraphics[width=0.47\linewidth]{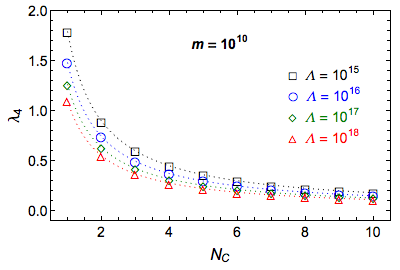}
   \includegraphics[width=0.47\linewidth]{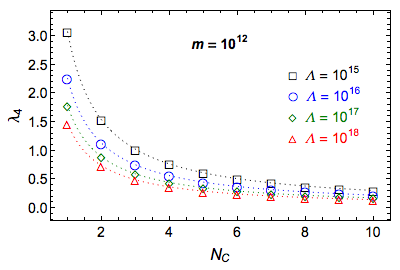}
    \includegraphics[width=0.47\linewidth]{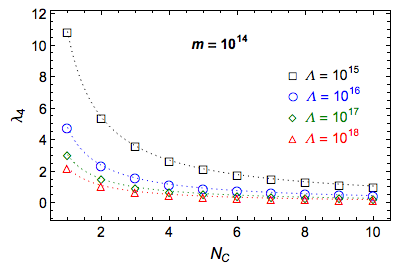}
 \caption{The self-coupling dependence on $N_{c}$ with the 4-momentum cutoff scheme $(\lambda := \lambda_{4})$ for different values of the NJL parameters $m$ and $\Lambda$, based on the 4MCS case of Eq.(\ref{scheme}).}
 \label{fig3}
\end{center}
\end{figure*}
Here we interpret $\Lambda$ as the scale of inflation. Commonly, it is expected to be the GUT energy scale, $O(10^{16})$\,GeV. Our predictions of the dependence of the self-coupling constant $\lambda$ on $N_{c}$ for the 3- and 4-momentum cutoff schemes for different values of $m$ and $\Lambda$ can be illustrated in Fig.\ref{fig2} (for the 3MCS) and in Fig.\ref{fig3} (for the 4MCS), respectively. From these figures one can see that when the dynamical mass $m$ is much less than the cutoff scale, there is no much difference using either 3MCS or 4MCS as it can be seen from Eq. (\ref{3or4}), while the values of $\lambda$ become significantly different when the dynamical mass $m$ is close to the cutoff scale. Taking the cases with $m = 10^{14}$GeV as an example, we find that $\lambda \approx 7.5$ for $N_c =2, \Lambda = 10^{15}$GeV in the 3MCS while $\lambda \approx 5.5$ for $N_c =2, \Lambda = 10^{15}$GeV in the 4MCS. Therefore one should be aware of the momentum cutoff scheme when the dynamical mass is not very far below the cutoff scale. Also it is worth noting that when $N_{c}$ is much greater than unity $(N_{c}\gg 1)$ the self-coupling evolves to zero in both schemes.

\subsection{Composite dark matter (CD) from the NJL}

In the previous section we have shown that the composite scalar $\varphi$ can play the role of inflaton and hence we inherently obtain a composite inflation model via an NJL formulation. As to this point our model is similar to other composite models (e.g. Ref.\cite{Inagaki:2015eza,Inagaki:2016vkf}), however,  the key points of our model are: First, we keep the composite pseudo-scalars as candidates for dark matter; Second, the composite scalar does NOT have to be the Higgs boson and hence, the dominating Yukawa-type coupling comes from the NJL four-fermion interaction, not the Yukawa coupling of the Higgs boson and the top quark in the Standard Model as in Ref. \cite{Clark09}. Let us recall the action given in Eq. (\ref{NJLcurvedxi1}) and add the Higgs sector 
\begin{eqnarray}
{\cal S}_{\textrm{\tiny{CD}}} &=& \int d^4x \sqrt{-g} \, \bigg[ - \frac{1}{2} M^{2}_{\textrm{\tiny{PL}}} R + \frac{1}{2}g^{\mu\nu}\partial_{\mu}\varphi\partial_{\nu}\varphi +\frac{1}{2}g^{\mu\nu}\partial_{\mu}S\partial_{\nu}S - \frac{\xi_s \, R}{2}\,  \left(\varphi^{2}+S^{2} - \frac{v^2}{2}\right) \cr &&\quad\quad\quad\quad\quad\quad - \frac{1}{2} m^{2}_{\varphi} \varphi^{2} - \frac{1}{2} m^{2}_{S} S^{2}-\frac{1}{2}\lambda\left(\varphi^{2}+S^{2}\right)^{2} + g^{\mu\nu}\partial_{\mu} H^{\dagger}\partial_{\nu} H \cr &&\quad\quad\quad\quad\quad\quad - \xi_h \,R\, H^{\dagger} H   -\lambda_h \left( H^{\dagger} H - \frac{v^2_0}{2} \right)^2 - \frac{1}{2}\kappa (\varphi^2 + S^2) H^{\dagger} H + \bar{\mathcal{L}}_{\text{\tiny{SM}}} \,\bigg], \label{Ldm0}
\end{eqnarray}
where other terms in the Standard Model are included in $\bar{\mathcal{L}}_{\text{\tiny{SM}}}$. We have added a subscript $s$ for $\xi$, i.e. $\xi = \xi_s$, to distinguish from the non-minimal coupling of the Higgs field to gravity, $\xi_h$. Now we consider the physics below the inflation scale and how it is connected to the electroweak theory.  Integrating out the heavy field $\varphi$, we obtain
\begin{eqnarray}
{\cal S}_{\textrm{\tiny{eff}}} &=& \int d^4x \sqrt{-g} \, \bigg[ - \frac{1}{2} M^{2}_{\textrm{\tiny{PL}}} R  +\frac{1}{2}g^{\mu\nu}\partial_{\mu}S\partial_{\nu}S - \frac{1}{2} \xi_s \, R \,  S^{2} - \frac{1}{2} m^{2}_{S} S^{2} - \frac{1}{2}\lambda \,S^{4} \cr && \quad\quad\quad\quad\quad\quad+ g^{\mu\nu}\partial_{\mu} H^{\dagger}\partial_{\nu} H - \xi_h \,R\, H^{\dagger} H   -\lambda_h \left( H^{\dagger} H - \frac{v^2_0}{2} \right)^2 \cr &&\quad\quad\quad\quad\quad\quad- \frac{1}{2} \, \kappa S^2 \, H^{\dagger} H + \bar{\mathcal{L}}_{\text{\tiny{SM}}} + \mathcal{O} \left(\frac{1}{m_{\varphi}^2}\right) \,\bigg], \label{Ldm}
\end{eqnarray}
where the term $\mathcal{O} (1/m_{\varphi}^2) $ includes interactions suppressed by a factor of $m_{\varphi}^2$. From (\ref{Ldm}) it is easy to see that the ``dark pion" field $S$ naturally appears as one of the messengers between the physics at the inflation scale and the physics at the electroweak scale. The Higgs field can be considered as another messenger, as it couples to both the inflaton $\varphi$ and other Standard Model fields. To describe the physics from high scales to low scales we need to study the renormalization group equations (RGEs) of the physical parameters in our model. At the one-loop level, the RGEs of the coupling ($\lambda_h, \kappa, \lambda, \xi_s, \xi_h, g_1, g_2, g_3, y_t$) ($g_1, g_2, g_3$ are gauge couplings of the Standard Model and $y_t$ is the Yukawa coupling of the top quark) \cite{Clark09, Lerner:2009xg}
\begin{eqnarray} \label{RGE}
(4\pi)^2 \frac{d g_1 }{d t}  &=& \frac{1}{12} g_1^3 (81 + c_h), \cr
(4\pi)^2 \frac{d g_2 }{d t}  &=& - \frac{1}{12} g_2^3 (39 - c_h), \cr
(4\pi)^2 \frac{d g_3 }{d t}  &=& - 7 g_3^3, \cr
(4\pi)^2 \frac{d y_t }{d t}  &=& y_t \left[ \left( \frac{23}{6} + \frac{2}{3} c_h \right) y_t^2 - 8 g^2_3 -\frac{17}{12} g_1^2 - \frac{9}{4} g_2^2 \right], \cr
(4\pi)^2 \frac{d \lambda_h }{d t}  &=&  ( 6 + 18 c^2_h) \lambda_h^2 - 6 y_t^4 + \frac{3}{8} [ 2g_2^4 + (g_1^2 + g_2^2)^2] + (12 y_t^2 - 3 g_1^2 - 9 g_2^2 ) \lambda_h + 2 c^2_s \kappa^2,  \cr
(4\pi)^2 \frac{d\kappa}{d t}  &=& \kappa \left[4 c_h c_s \kappa + 6 (1+ c_h^2) \lambda_h + 6 c_s^2\lambda   - \frac{3}{2} (3 g_2^2 + g_1^2) + 6 y_t^2  \right] , \cr
(4\pi)^2 \frac{d \lambda}{d t}  &=& 18 c_s^2  \lambda^2 + \frac{1}{2} (3+ c_h^2) \kappa^2\,,
\end{eqnarray}
where $t \equiv \ln(\mu/m_t)$ and $\mu$ is the renormalization scale ($m_t \approx 173$ Gev is the top quark mass), and $c_h, ~c_s$ are the suppression factors for the Higgs and the $S$ field, respectively. They are brought in by the non-minimal coupling to the Ricci scalar and introduce a modification to the Higgs and the $S$ field propagators, respectively. To see how this happens one may use the effective action (\ref{Ldm}) to write the Einstein equation and the equations of motion for the Higgs and the $S$ field, respectively, and then combine them to identify the suppression factors (see e.g. \cite{Clark09}. Other approaches, say considering commutation relation in both Jordan and Einstein frames yields the same result \cite{Lerner:2009xg}).

To complete the RGEs we have to include the running of the non-minimal couplings $\xi_h$ and $\xi_s$, \cite{Clark09, Lerner:2009xg}
\begin{eqnarray} \label{RGE2}
(4\pi)^2 \frac{d \xi_s}{d t}  &=& \left(\xi_s + \frac{1}{6} \right) ~6 c_s  \lambda + \left(\xi_h + \frac{1}{6}\right) ( 3 + c_h^2) \kappa,  \cr
(4\pi)^2 \frac{d \xi_h}{d t}  &=& \left(\xi_h + \frac{1}{6} \right) \left[ 6 (1 + c_h^2) \lambda_h + 6  y_t^2 - \frac{3}{2} (3 g_2^2 + g_1^2) \right] + \left(\xi_s + \frac{1}{6}\right)  c_s \kappa\,.
\end{eqnarray} 
Interestingly, the behaviour of the scalar-gravitational coupling constant $\xi(t)$ of the class of gauge-Higgs-Yukawa models and the gauged Nambu-Jona-Lasinio (NJL) model was so far discussed in Refs.\cite{Geyer:1996kg,Buchbinder:1985ba}. In our case, inflation is driven by the composite scalar $\varphi$, not the Higgs field $H$, and correspondingly the suppression factors $c_h$ and $c_s$ are chosen to be 
\begin{equation}
c_h = 1, ~~~~c_s = \frac{1 + \xi_s S^2/M^{2}_{\textrm{\tiny{PL}}}}{1+ (1+ 6 \xi_s)  \, \xi_s S^2/M^{2}_{\textrm{\tiny{PL}}} }= \frac{1}{1+ 6 \xi_s \frac{1}{1+ M^{2}_{\textrm{\tiny{PL}}}/\xi_s S^2}} ,
\end{equation} 
respectively, and it is easy to see that 
\begin{equation}
c_s \approx 1, ~~\textrm{when}~~\xi_s S^2 \ll M^{2}_{\textrm{\tiny{PL}}}, ~~~~~c_s \approx \frac{1}{1 +  6 \xi_s}, ~~\textrm{when}~~\xi_s S^2 \gg M^{2}_{\textrm{\tiny{PL}}} ,
\end{equation}
hence we know that $c_s \approx 1$ at low scales and drops to zero at high scales, since from the inflation constraints $\xi_s \sim \mathcal{O} (10^4)$ for $\lambda \sim 0.5$ (see Fig. 3).  

A complete numerical study of the above RGEs will be done in a separate publication \cite{Channuie16}. Here we use these RGEs to estimate the range of the parameters. 
Let us start with the RGE of $\xi_s$,  
\begin{equation}
(4\pi)^2 \frac{d \xi_s}{d t}  = \left(\xi_s + \frac{1}{6} \right) ~6 c_s  \lambda + \left(\xi_h + \frac{1}{6}\right) 4 \kappa \approx 6 c_s \xi_s \lambda,
\end{equation}
where we have used the assumption that $\xi_s \gg \xi_h$ and $\xi_s \gg 1/6$. It can be rewritten as
\begin{equation}
\frac{d \ln \xi_s}{d t} \approx \frac{3}{8 \pi^2} c_s \lambda
\end{equation}
which leads to an approximate solution (with an initial value $\xi_s |_{t=0} = \xi_s^0$)
\begin{equation} \label{xi_s}
\xi_s (t) \approx \xi_s^0 ~e^{\int^t \frac{3}{8 \pi^2} c_s \lambda ~dt'}.
\end{equation}
Next we look at the RGE of $\lambda$. Since $\lambda$ and $\kappa$ are at the same order,
\begin{equation}
(4\pi)^2 \frac{d \lambda}{d t}  = 18 c_s^2  \lambda^2 + 2 \kappa^2 \approx  18 c_s^2  \lambda^2. 
\end{equation}
We use a step-function to describe the profile of the suppression factor $c_s$, i.e.
\begin{equation} \label{c_s}
c_s (t) = \begin{cases}
        ~~~1 & \text{when} ~t \leqslant t_s \\
        &\\
       ~~~0 & \text{when} ~t > t_s
        \end{cases}
\end{equation}
where $t_s$ is the scale indicating some new physics or a new phase for the scalar field $S$, e.g. a strongly coupled phase which reflects that it is getting closer to the composite scale. By asking the new composite sector to be responsible for this new physics or new phase, instead of the Higgs sector, our model is less constrained than the Higgs inflation models and naturally contains a strongly coupled phase.  This is similar to the study on extending the Standard Model with the fourth generation \cite{Hung:2010xh, Hung:2009hy}, where the constituent fermions are the fourth generation quarks and leptons and their bound states yield extra scalars, but the Higgs field is unconstrained and the strongly coupled phase only happens to the fourth generation fermions. 
Of course one can smooth out the step-function behavior of the suppression factor $c_s$ by using $c_s(t) = 1/ [1 + e^{\alpha (t-t_s)}], ~(\alpha > 0)$.  However, Eq. (\ref{c_s}) is good enough for the estimate purpose. 
With an initial value $\lambda |_{t=0} = \lambda_0$, the above equation gives
\begin{equation}
\lambda (t) \approx \frac{\lambda_0}{1 - \frac{9 \lambda_0}{8 \pi^2} t }, ~~~~~\textrm{for}~~t \leqslant t_s
\end{equation}
Plugging into Eq. (\ref{xi_s}) we obtain
\begin{equation}
\xi_s (t) \approx \xi_s^0 ~e^{ -\frac{1}{3} \ln \left( 1 - \frac{9 \lambda_0}{8 \pi^2} t \right)} = \xi_s^0  \left( 1 - \frac{9 \lambda_0}{8 \pi^2} t \right)^{-\frac{1}{3}},  ~~~~~\textrm{for}~~t \leqslant t_s
\end{equation}
Taking $\lambda_0 = 0.2$ and $t_s = 27$ (in comparison with \cite{Clark09, Lerner:2009xg}), we find the value of $\xi_s$ at $t = t_s$ is 
\begin{equation}
\xi_s |_{t=27} \approx 1. 4 \, \xi_s^0 
\end{equation}
Now we consider the RGE for $\xi_{h}$
\begin{equation}
(4\pi)^2 \frac{d \xi_h}{d t}  = \left(\xi_h + \frac{1}{6} \right) \left[ 12 \lambda_h + 6  y_t^2 - \frac{3}{2} (3 g_2^2 + g_1^2) \right] + \left(\xi_s + \frac{1}{6}\right)  c_s \kappa  \approx \xi_s c_s \kappa,
\end{equation}
which leads to an approximate solution 
\begin{equation}
\xi_h = \frac{1}{16 \pi^2} \int_0^t \xi_s(t') c_s \kappa(t') dt'
\end{equation}
with an initial condition that $\xi_h |_{t=0} =0$. 
The RG-running of the coupling $\kappa$ is more difficult to estimate since the $\beta$-function of $\kappa$ contains terms with different signs and those terms are at the same order $\mathcal{O}(1)$ (with coefficients multiplied). Here we assume that $\kappa$ is a slow-changing parameter and use its average, $\bar{\kappa}$ to replace $\kappa(t')$ in the above integral, 
\begin{equation}
\xi_h \approx \frac{\xi_s^0  \bar{\kappa}}{18 \lambda_0} \left[ 1 - \left( 1 - \frac{9 \lambda_0}{ 8 \pi^2} t \right)^{\frac{2}{3}} \right],  ~~~~~\textrm{for}~~t \leqslant t_s.
\end{equation}
Therefore we find the ratio
\begin{equation}
\frac{\xi_h}{\xi_s} \approx \frac{\bar{\kappa}}{18 \lambda_0}  \left[ \left( 1 - \frac{9 \lambda_0}{ 8 \pi^2} t \right)^{\frac{1}{3}} -  \left( 1 - \frac{9 \lambda_0}{ 8 \pi^2} t \right) \right],~~~~~\textrm{for}~~t \leqslant t_s.
\end{equation}
For $\bar{\kappa} \approx \lambda_0 = 0.2$ and $t_s = 27$, we obtain that $\xi_h \approx 0.019 \,\xi_s$. Therefore we find that even we impose the initial condition that $ \xi_h |_{t=0} =0$, it will evolve to some non-vanishing value at high-scales, although still being dominated by the $\xi_s$ coupling. It is important that the RG-running of $\xi_h$ does not spoil our assumption on the dominance of the composite inflation over the Higgs inflation. 

\section*{CD Relic abundance}
Nowadays the physical properties of the Higgs sector in the standard model are very accurate. This in general allows us to couple any additional sector to the Higgs one in a unique way. Here in the present work we examine physical parameters/constraints of the model of DM by coupling the field $S$ to the Higgs sector. What we are going to discuss below is similar to the Higgs-portal paradigms, see Ref.\cite{Han:2015hda} for example, which contain the coupling constant between two Higgs bosons and two new scalars. This is what we have in Eq.(\ref{Ldm}). We find from Eq.(\ref{Ldm}) the physical mass of the dark matter scalar $S$:
\begin{equation}
M_{s} \approx \sqrt{m^{2}_{s} + \kappa v^{2}_{0}/2}\,\,{\rm with}\,\,v_{0}=246\,{\rm GeV}\,. \label{dmass}
\end{equation}
In order to determine the relic density of $M_{s}$ in the vicinity of the resonance at center of mass energy $s = 4M^{2}_{s}$, it is essential to figure out the thermally averaged annihilation cross section as a function of $x=M_{s}/T$ given by \cite{Gondolo:1990dk,Cline:2013gha} 
\begin{equation}
\left<\sigma v_{\rm rel}\right> (x)= \frac{x}{16M^{5}_{S}K^{2}_{2}(x)}\int^{\infty}_{4M^{2}_{s}}s^{3/2}\,\sigma v_{\rm rel}\, \sqrt{1-\frac{4M^{2}_{s}}{s}}K_{1}\left(\frac{\sqrt{s}}{M_{s}}x\right)ds\, \label{relic}
\end{equation}
in terms of the modified Bessel functions of the second kind $K_{1,2}$, and to solve the Boltzmann equation for the relic abundance \cite{Lee:1977ua}. In the present work, we will consider the tree-level processes contributing to $S$ annihilation via $SS\rightarrow \{\,f\bar{f},\,WW,\,ZZ,\,hh\,\}$ with $f$ being a standard model fermion and estimate $\left<\sigma v_{\rm rel}\right>$ by the center-of-mass cross section for non-relativistic $S$ annihilation. The corresponding $\left<\sigma v_{\rm rel}\right>$ read \cite{Lerner:2009xg,Cline:2012hg}:
\begin{eqnarray} \label{RGEto}
\left<\sigma v_{\rm rel}\right>_{f\bar{f}} &=& \frac{M^{2}_{W}}{\pi g^{2}}\frac{\kappa^{2}\left(M_{f}/v_{0}\right)^{2}}{\left(\left(4M^{2}_{s}-M^{2}_{h}\right)^{2} + M^{2}_{h}\Gamma^{2}_{h}\right)}\left(1 - \frac{M^{2}_{f}}{M^{2}_{s}}\right)^{3/2}\,,\cr
\left<\sigma v_{\rm rel}\right>_{WW}  &=& 2\left(1 + \frac{1}{2}\left(1 - \frac{2M^{2}_{s}}{M^{2}_{W}}\right)^{2}\right)\left(1 - \frac{M^{2}_{W}}{M^{2}_{s}}\right)^{1/2}\frac{\kappa^{2}M^{4}_{W}}{8\pi M^{2}_{s}\left(\left(4M^{2}_{s}-M^{2}_{h}\right)^{2} + M^{2}_{h}\Gamma^{2}_{h}\right)}\,, \cr
\left<\sigma v_{\rm rel}\right>_{ZZ}   &=&  2\left(1 + \frac{1}{2}\left(1 - \frac{2M^{2}_{s}}{M^{2}_{Z}}\right)^{2}\right)\left(1 - \frac{M^{2}_{Z}}{M^{2}_{s}}\right)^{1/2}\frac{\kappa^{2}M^{4}_{Z}}{16\pi M^{2}_{s}\left(\left(4M^{2}_{s}-M^{2}_{h}\right)^{2} + M^{2}_{h}\Gamma^{2}_{h}\right)}\,,  \cr
\left<\sigma v_{\rm rel}\right>_{hh}   &=& \frac{\kappa^{2}}{64\pi M^{2}_{s}}\left[1 + \frac{3M^{2}_{h}}{(4M^{2}_{s} - M^{2}_{h})} + \frac{2\kappa v^{2}_{0}}{M^{2}_{h} - 2M^{2}_{s}}\right]^{2}\left(1 - \frac{M^{2}_{h}}{M^{2}_{s}}\right)^{1/2}\,,
\end{eqnarray}
\begin{figure*}
\begin{center}
 \includegraphics[width=0.48\linewidth]{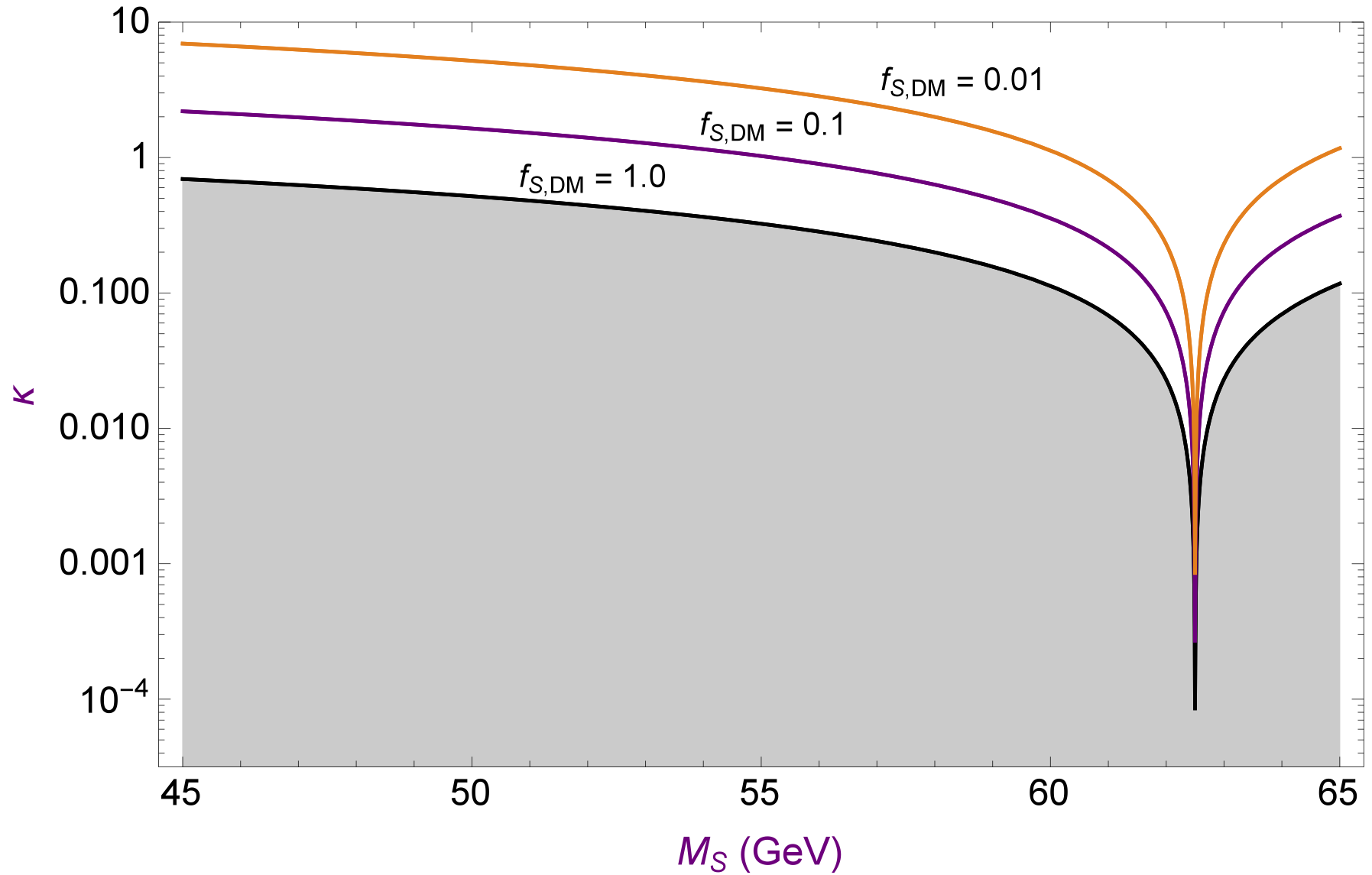}
  \includegraphics[width=0.48\linewidth]{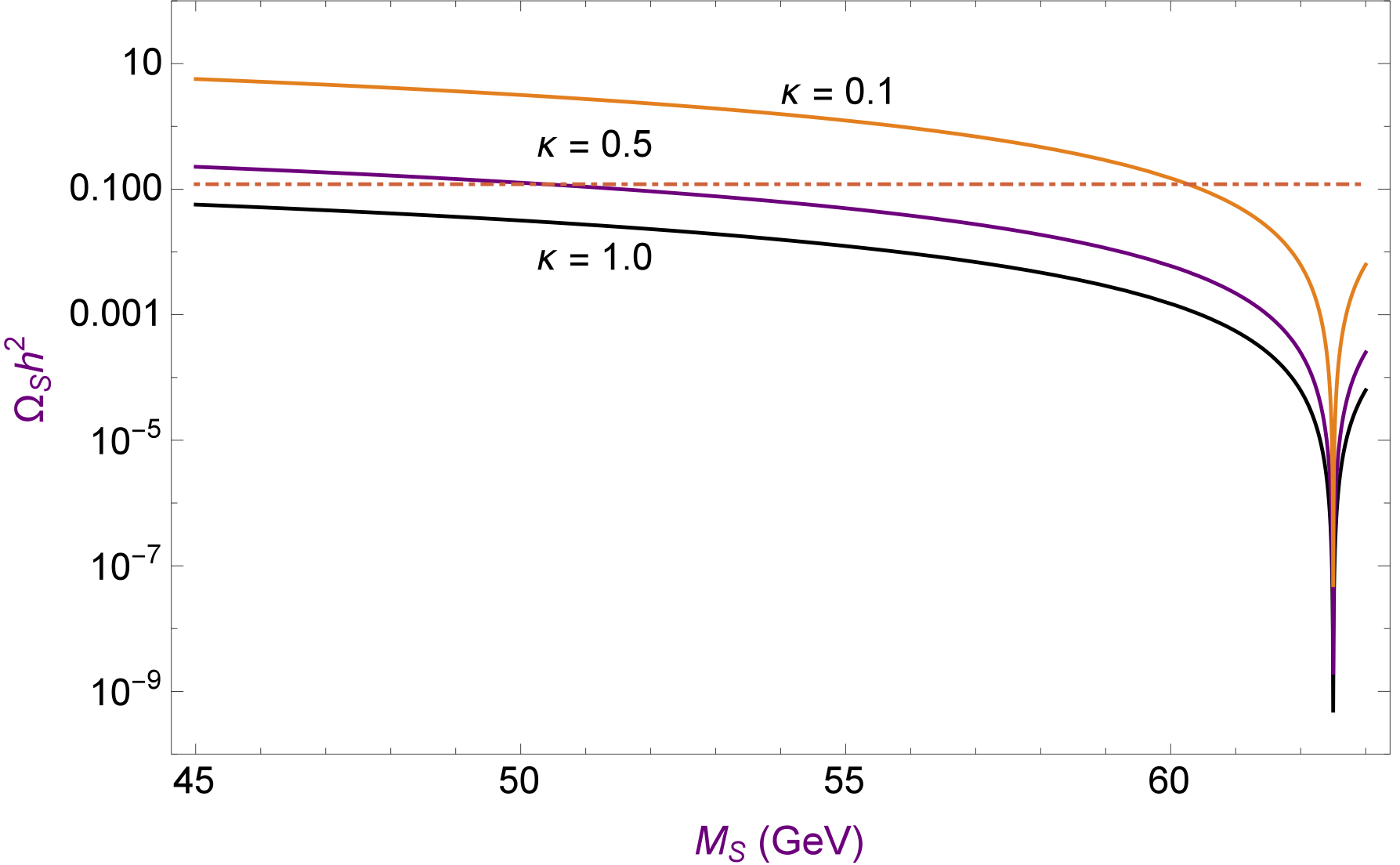}
 \caption{In case of light DM, we plot in left-panel the two-parameters plane of $(\kappa,\,M_{s})$ implemented by Eq.(\ref{fract}) using only the thermally averaged annihilation cross section $\left<\sigma v_{\rm rel}\right>_{SS\rightarrow b{\bar b}}$ and $x_{f}=20$ for three different fractions $f_{S,{\rm DM}}$. A shaded-lower region is ruled out since they produce more than the observed relic density of dark matter. Right panel: The scalar singlet DM abundance $\Omega_{S}{\rm h}^{2}$ as a function of the DM mass $M_{s}$ in unit of GeV by imposing three different values of $\kappa$ with $x_{f}=20$. The horizontal dotdashed-line represents the observed relic abundance of the DM, $\Omega_{\rm DM}{\rm h}^{2} = 0.1199$.}
 \label{fig6}
\end{center}
\end{figure*}
where the first three contributions proceed via $s$-channel Higgs exchange, while the last one comes from the $s$-channel Higgs exchange interaction and a $t$- and $u$-channel $S$ exchange interaction, $M_{f}$ is the SM fermion mass, and $\Gamma_{h}$ is the total Higgs decay width: $\Gamma_{h} = 6.1^{+7.7}_{-2.9}\,{\rm MeV}$ \cite{Barger:2012hv}. The freezout value $x=x_{f}$ can be iteratively determined in this case using the relation \cite{Aravind:2015xst} 
\begin{equation}
x_{f} \equiv \frac{M_{s}}{T_{f}}= \ln\left(\frac{3M_{\textrm{\tiny{PL}}}}{4\pi^{2}}\sqrt{\frac{5M^{2}_{S}}{\pi g_{*}x_{f}}}\left<\sigma v_{\rm rel}\right> (x)\right)\,, \label{ssrelic}
\end{equation}
where $g_{*}$ is the number of relativistic degrees of freedom at the freeze out temperature. With all mass dimensions expressed in GeV, the relic density as a function of the non-relativistic annihilation cross section is of the form \cite{Campbell:2016llw} 
\begin{equation}
\Omega_{S}{\rm h}^{2} \approx \,1.65 \times 10^{-10}\,x_{f}\,\frac{\left({\rm GeV^{2}}\right)}{\left<\sigma v_{\rm rel}\right> (x_{f})}\,, \label{ssrelicd}
\end{equation}
where ${\rm h}$ is related to the Hubble parameter at the present time $H_{0}$ via ${\rm h} := H_{0}/(100\,\frac{\rm km}{\rm s\,Mpc})\approx 0.7$, $\left<\sigma v_{\rm rel}\right>$ stands for the thermally-averaged annihilation cross section (times relative velocity). In order to reproduce the relic density in agreement with the observed value, the composite DM $S$ must annihilate at early time with a suitable cross section.
\begin{figure*}
\begin{center}
 \includegraphics[width=0.48\linewidth]{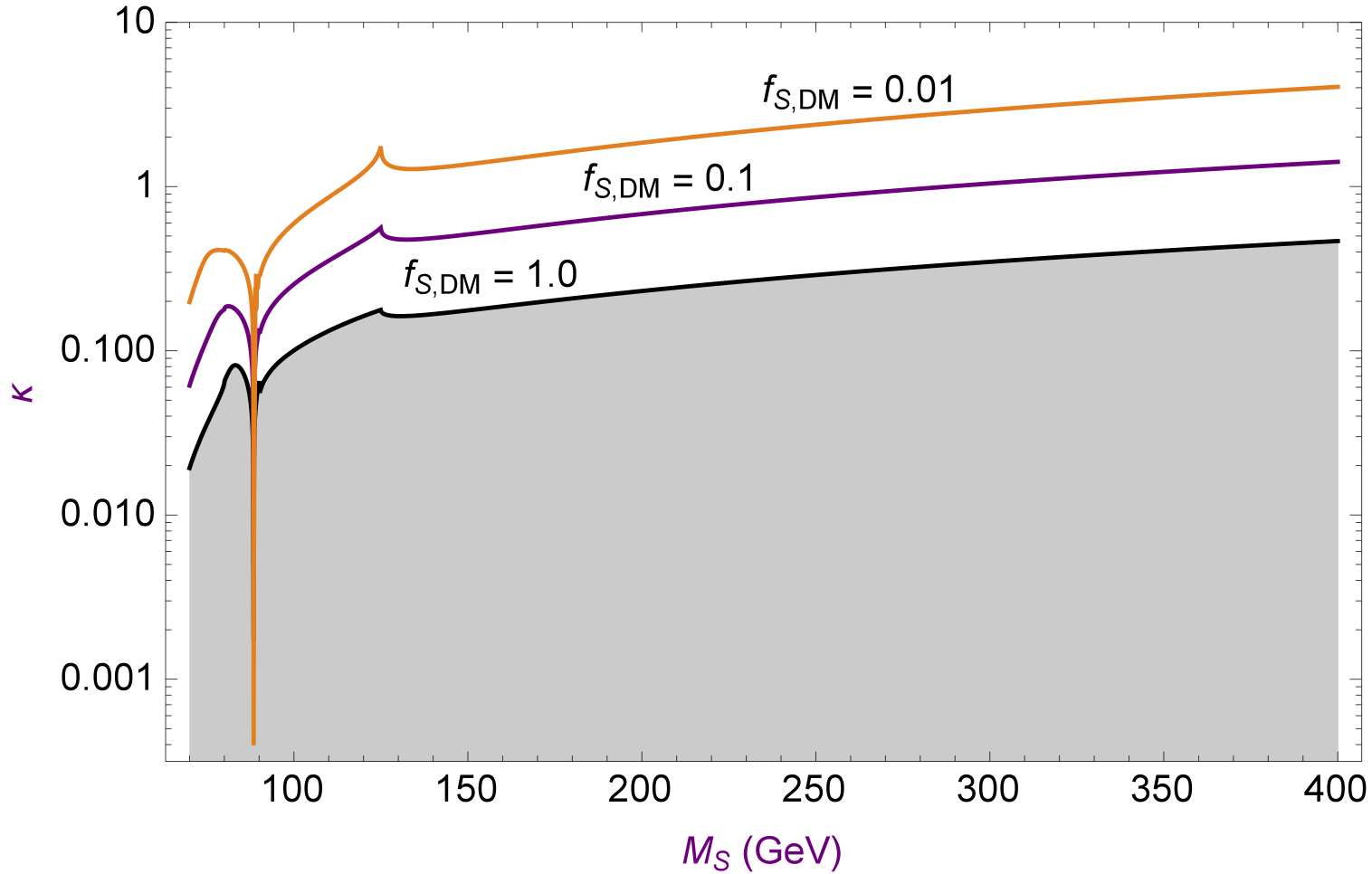}
  \includegraphics[width=0.48\linewidth]{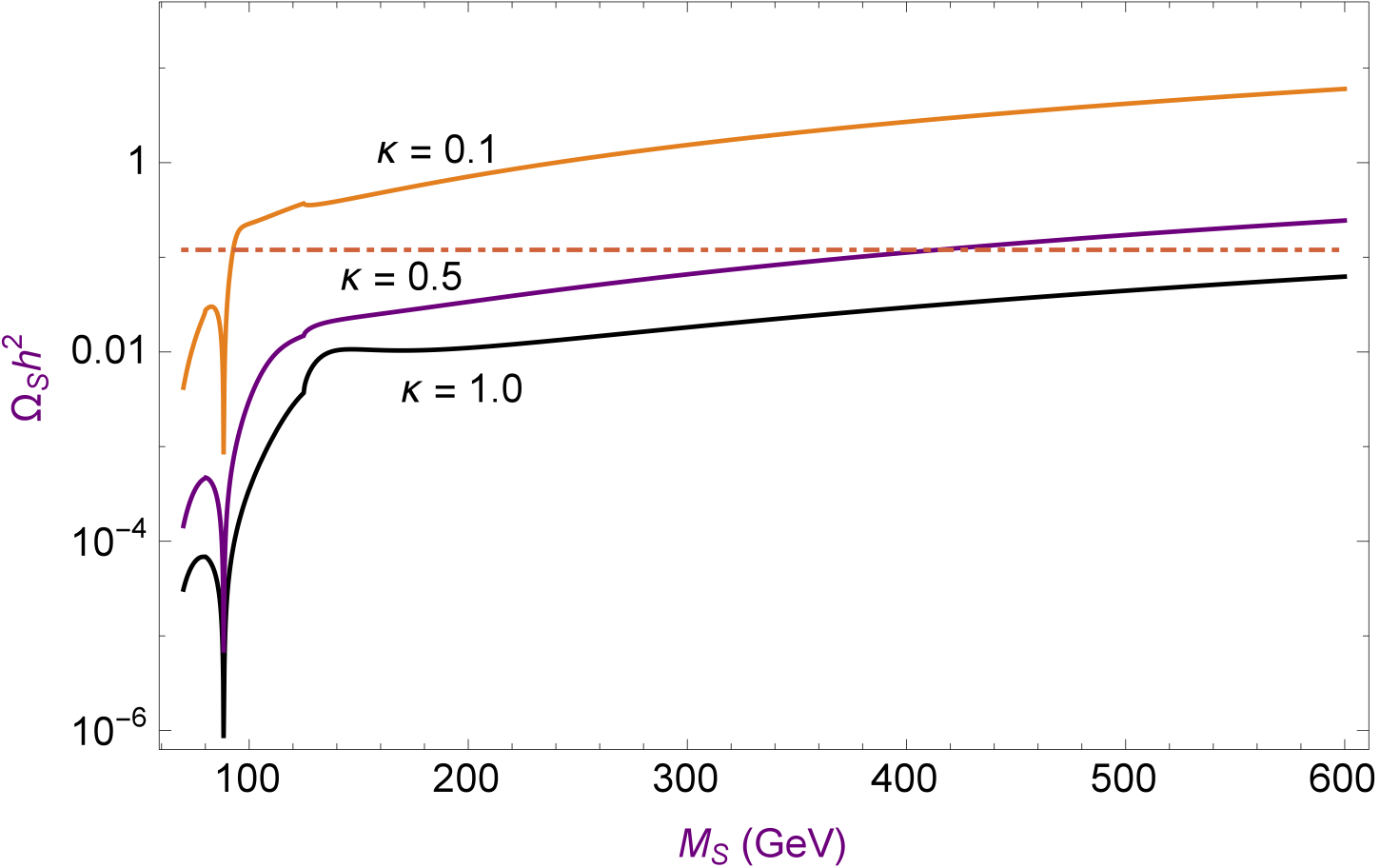}
 \caption{In the case of heavy DM, the plot in the left panel shows the two-parameters plane of $(\kappa,\,M_{s})$ implemented by Eq.(\ref{fract}) using the thermally averaged annihilation cross sections $\left<\sigma v_{\rm rel}\right>_{b{\bar b},\,WW,\,ZZ,\,hh}$ and $x_{f}=20$ for three different values of $f_{S,{\rm DM}}$. A shaded-lower region is ruled out since they produce more than the observed relic density of dark matter. The right panel: The scalar singlet DM abundance $\Omega_{S}{\rm h}^{2}$ as a function of the DM mass $M_{s}$ in unit of GeV by imposing three different values of $\kappa$ with $x_{f}=20$. The horizontal dotdashed-line represents the observed relic abundance of the DM, $\Omega_{\rm DM}{\rm h}^{2} = 0.1199$.}
 \label{fig7}
\end{center}
\end{figure*}
Ref.\,\cite{Ade:2013zuv} reported the present relic density of dark matter to be $\Omega_{\rm DM}{\rm h}^{2} = 0.1199\pm 0.0027$. It was mentioned in Ref.\,\cite{Lerner:2009xg} that at the tree level process the cross sections for real and complex $S$ are the same. In order to quantify the DM abundance, we may define the fraction $f_{S,{\rm DM}}$:
\begin{equation}
f_{S,{\rm DM}}\equiv \frac{\Omega_{S}{\rm h}^{2}}{0.1199} \approx \,14 \times 10^{-10}\,x_{f}\,\frac{\left({\rm GeV^{2}}\right)}{\left<\sigma v_{\rm rel}\right> (x_{f})}\,, \label{fract}
\end{equation}
If $f_{S,{\rm DM}}<1$, then the relic density of $S$ is suppressed relative to the observed value. From Eq.(\ref{Ldm}), we have the $SSh$ coupling term which mediates $SS$ interactions with pairs of SM-particles through the light Higgs pole, and in general Higgs decays $h\rightarrow SS$ are also allowed. We will start quantifying the DM relic abundance by first supposing the DM is light such that $M_{s} < M_{h}/2$. In this low mass case, the decay $h\rightarrow SS$ is kinematically allowed, and contributes to the invisible width $\Gamma_{\rm inv}$ of the Higgs boson. The cross section in this case is dominated by the dark matter annihilation process to a pair of bottom quarks $SS\rightarrow b{\bar b}$ (which the Higgs mediates the interaction) with a branding ratio around $60\%$. When assuming that a composite scalar field $S$ is responsible for the dark matter density, then we obtain the relationship between $\kappa$ and $M_{s}$ illustrated in Fig.(\ref{fig6}).

In Fig.(\ref{fig6}), in the left panel, we plot the plane of $M_{s}$ and the coupling $\kappa$. We also display them in the low mass limit over the range of DM mass values ($45\,\,{\rm GeV}\leq M_{s}\leq 65\,\,{\rm GeV}$), and in the region $M_{s}=M_{h}/2$ where annihilation is resonantly enhanced. We expect when including the Higgs invisible width that below $M_{h}/2$ a small triangle in the $\kappa-M_{s}$ plane will survive (see for example Ref.\cite{Cline:2013gha}). In the right panel of Fig.(\ref{fig6}), we diplay the relic abundance $\Omega_{S}{\rm h}^{2}$ as a function of the DM mass, $M_{s}$, for different values of the coupling. We find that the required DM abundance is archived for values of $M_{s}\sim 61\,{\rm GeV}$ for $\kappa = 0.1$. However, the CD mass can be lighter when the coupling is getting bigger. 

However, in the case of a heavy DM, i.e. $M_{s}> M_{W,\,Z}$ (or $>M_{h}$), the contributions from $SS\rightarrow \{\,WW,\,ZZ,\,hh\,\}$ are allowed for quantifying the DM relic abundance. In this case, we also obtain the relationship between $\kappa$ and $M_{s}$ illustrated in Fig.(\ref{fig7}). In Fig.(\ref{fig7}), in the left panel, we plot the plane of $M_{s}$ and the coupling $\kappa$. We also display them in the heavy mass limit over the broad range of DM mass values, and in the region $M_{s} \approx M_{Z}$ where annihilation is resonantly enhanced. Moreover, in the region above $M_{h}/2$, the authors of Ref.\cite{Cline:2013gha} show that the relic density constrains the coupling as a function of the DM mass which can be approximately describe by the dependence $\log_{10}\kappa > -3.63 + 1.04 \log_{10}(M_{s}/{\rm GeV})$. In the right panel of Fig.(\ref{fig7}), we also diplay the relic abundance $\Omega_{S}{\rm h}^{2}$ as a function of the DM mass, $M_{s}$, for different values of the coupling. We find that the required CD relic abundance is archived for values of $M_{s}\sim 410\,{\rm GeV}$ for $\kappa = 0.5$. In contradiction to the light mass case, however, the CD mass in this case can be heavier when the coupling is getting larger. For more accurate investigation, we acknowledge, for instance, Refs.\cite{Cline:2013gha,Campbell:2016llw,Guo:2010hq,Han:2015hda} and references therein.
\begin{figure*}
\begin{center}
 \includegraphics[width=0.48\linewidth]{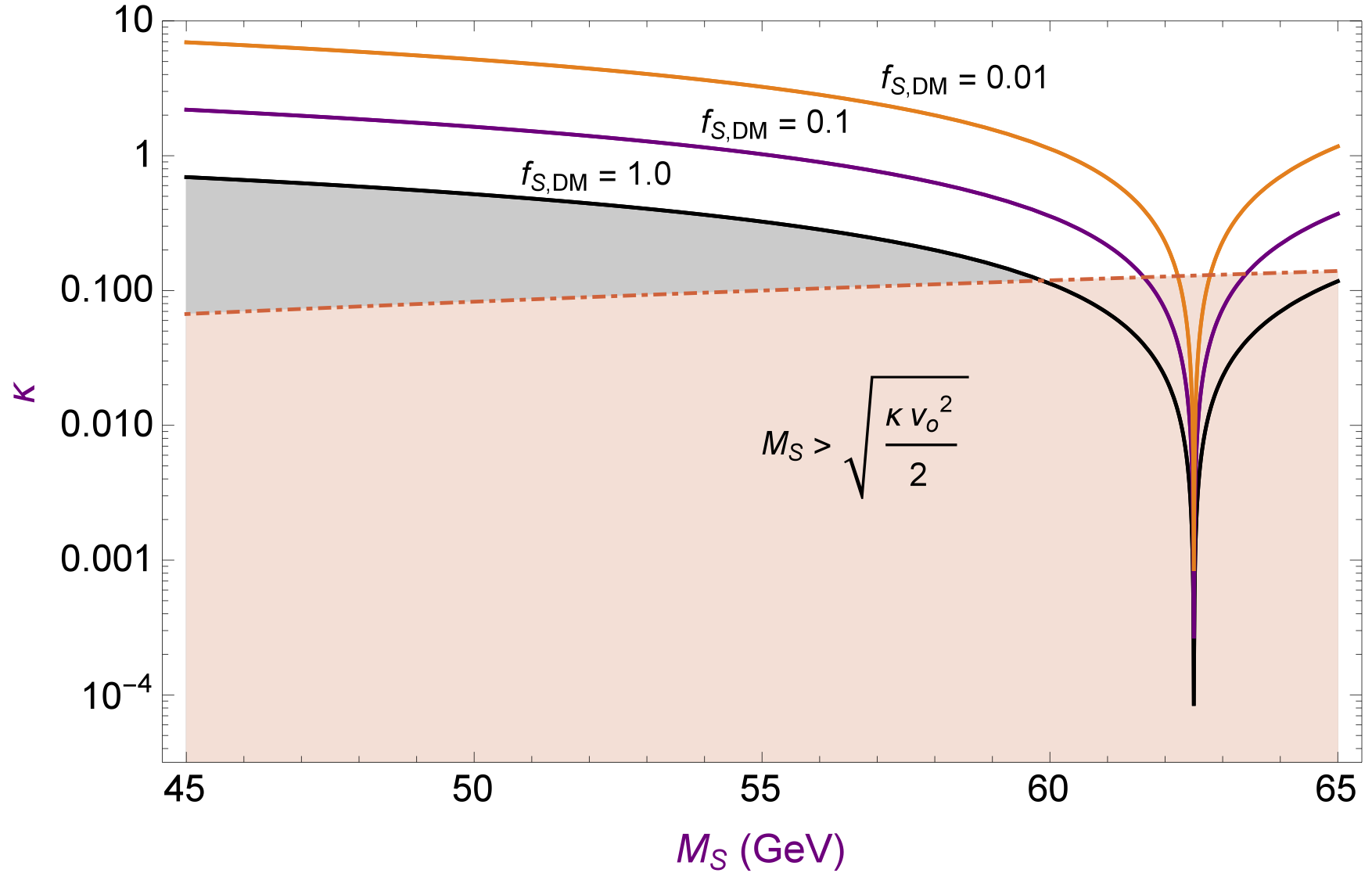}
  \includegraphics[width=0.48\linewidth]{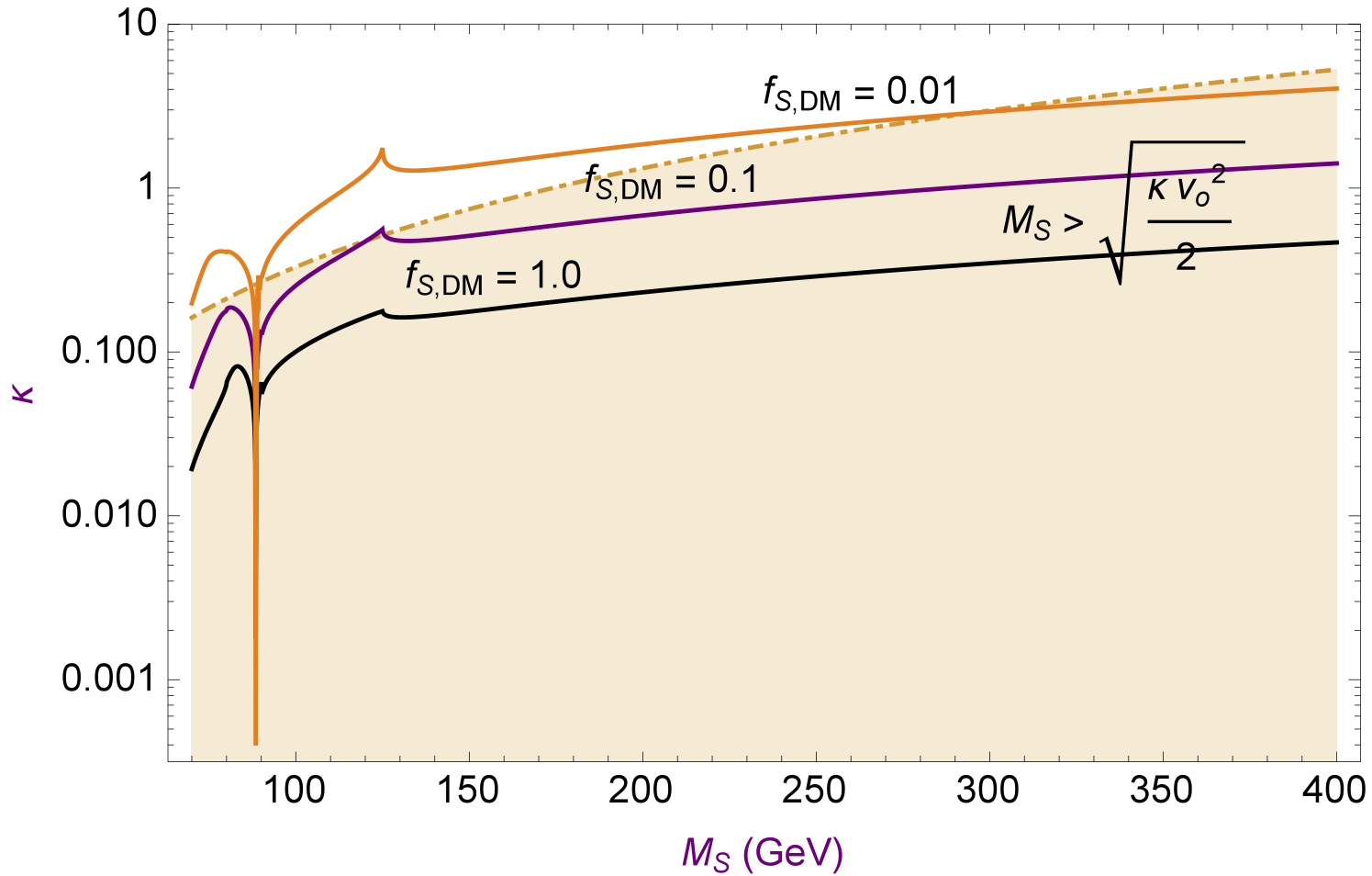}
 \caption{The plot shows the two-parameters plane of $(\kappa,\,M_{s})$ implemented by Eq.(\ref{fract}) using the thermally averaged annihilation cross sections: left panel for the light CD and right panel for the heavy one. A shaded region just below a dot-dashed line satisfies a condition $M_{s} > (\kappa v^{2}_{0}/2)^{1/2}$ in which the parameters are allowed.}
 \label{fig8}
\end{center}
\end{figure*}
Moreover, Eq.(\ref{dmass}) allows us to visualize some more constraints on the parameter spaces. Since $m^{2}_{s}$ is positive, Eq.(\ref{dmass}) suggests a constraint $M_{s}>(\kappa v^{2}_{0}/2)^{1/2}$. As a result, in Fig.\ref{fig8}, we have another curve of $M_{s} = (\kappa v^{2}_{0}/2)^{1/2}$ below which the parameters are allowed. We can be even more concrete. In order to satisfy the constraints, from Fig.(\ref{fig8}) in case of the light CD, we could have $\kappa\lesssim 0.1$ (together with the regions above $f_{S,{\rm DM}}=1.0$ line) and the CD mass in the vicinity of the resonance will survive, while in case of the heavy one the allowed couplings $\kappa$ stay in between the $f_{S,{\rm DM}}=1.0$ line and the dot-dashed one.

We finalise this section by relating the CD parameters to the NJL ones and supporting for example that $\Lambda\gg m$ and $\Lambda_{3}\gg m$. In this case, the two different momentum-cutoff schemes approach the same result: $I_2 \approx \frac{1}{8 \pi^2} \ln\left(\frac{\Lambda}{m} \right)$. 
Interestingly, since $\lambda$ is related to the NJL parameters as
\begin{equation} \label{NcI2}
N_{c}I_{2} \approx \frac{1}{4\lambda},
\end{equation}
we can now use the above result to incorporate the CD parameter set with the NJL ones. From Eq.(\ref{NJLset2}), we obtain the relation
\begin{eqnarray}
\left(\frac{m^{2}_{s}}{M^{2}_{G}} \right) \approx \sqrt{\lambda}\left(\frac{m_0}{m}\right). \label{NJLset3}
\end{eqnarray}
Let's take for example $m\sim 10^{12}\,{\rm GeV}$ and $\lambda \approx 0.5$. We obtain for the light CD and the heavy one, respectively
\begin{equation} \label{approx}
 m_{0} M^{2}_{G} \approx \begin{cases}
        \,\, 1.2 \times 10^{15} ~{\rm GeV^{3}}, & {\rm for}\,\,m_{s} = \sqrt{M^{2}_{s} - \frac{\kappa v^{2}_{0}}{2}} \approx 29\,{\rm GeV}\,,\\
        &\\
        \,\, 2.2 \times 10^{17} ~{\rm GeV^{3}}, & {\rm for}\,\,m_{s} = \sqrt{M^{2}_{s} - \frac{\kappa v^{2}_{0}}{2}} \approx 391\,{\rm GeV},
        \end{cases}
\end{equation}
where we have used in case of light CD $M_{s}\sim 61\,{\rm GeV},\,\kappa=0.1$ and in case of heavy one $M_{s}\sim 410\,{\rm GeV},\,\kappa=0.5$. These values can be obtioned from Fig.(\ref{fig6}) and Fig.(\ref{fig7}). Hopefully, the measurement of values of $m_{0}M^{2}_{G}$ may in principle turn out to be critical in testifying our model. Note that a large value for $M_G$ can drive the value of $m_0$ to be very small, i.e. the ``current" mass of the composite fermions is extremely small compared with their dynamical mass.

\section{conclusion}

In this work, we presented a unified description of inflation and dark matter in the context of the effective NJL model. We also demonstrated how an NJL effective potential emerges and proposed a cosmological scenario that unifies comic inflation and dark matter to a single framework. On the one hand, we showed that the scalar channel of the NJL model with a non-minimal coupling to gravity plays a role of the composite inflaton (CI). For model of inflation, we computed the inflationary parameters and confront them with recent Planck 2015 data. We discovered that the predictions of the model are in excellent agreement with the Planck analysis. 

We presented in our model a simple connection of physics from the high scales to low scales via renormalization group equations (RGEs) of the physical parameters and use them to estimate the range of relevant parameters. On the other hand, the pseudoscalar channel can be assigned as a candidate for composite dark matter (CD). For model of dark matter, we coupled the pseudoscalar to the Higgs sector of the standard model with the coupling strength $\kappa$ and estimate its thermally-averaged relic abundance. We discovered that the CD mass is strongly sensitive to the coupling $\kappa$. We found in case of light CD, $M_{s}<M_{h}/2$, that the required relic abundance is archived for value of its mass $M_{s} \sim 61\,{\rm GeV}$ for $\kappa=0.1$. However, in this case the CD mass can be lighter when the coupling is getting larger. Moreover, in case of heavy CD, $M_{s}> M_{W,\,Z}$ (or $>M_{h}$), the required relic abundance is archived for values of the CD mass $M_{s}\sim 410\,{\rm GeV}$ for $\kappa = 0.5$. In contradiction to the light mass case, however, the CD mass in this case can even be heavier when the coupling is getting larger.

There are some limitations in the present work --- for example, the effective potential we used for computing the inflationary parameters does not include the RG-improved part as in Refs.\cite{Inagaki:2015eza,Inagaki:2016vkf}; We have not considered any other channels than the scalar and pseudoscalar channels; The NJL description itself can be more general, say including more flavors, or including some gauge fields so it becomes a gauged NJL model. Regarding our present work, it is possible to extend this study to account of two-loop effects of inflationary model, see Ref.\cite{Inagaki:2015fva}. Moreover, a determinant term can be added to the NJL action so one can take anomaly into account \cite{Hatsuda, Klevansky}; Also one should complete the RGEs for all scales and solve them numerically. We hope to address these issues with future investigations. Moreover, regarding this single framework, another crucial issue for successful models of inflation is the (pre)reheating mechanism. We plan to investigate this mechanism, within our framework, by following one of the very recent examinations on the (pre)reheating mechanism underlying a composite inflationary scenario \cite{Channuie:2016xmq}.

More recently, direct searches for DM by the LUX and PandaX-II Collaborations \cite{Akerib:2016vxi,Tan:2016zwf} implementing xenon-based detectors have recently come up with the most stringent limits to date on the elastic scattering of DM off nucleons, see Ref.\cite{He:2016} for the very recent analysis. In the simplest model, SM+DM, which is the standard model plus a real scalar singlet (darkon) acting as the DM candidate, the LUX and PandaX-II limits rule out DM masses from $5\,{\rm GeV}$ to about $330\,{\rm GeV}$, except a small triangle around the resonant point at half of the Higgs mass. Therefore, it is reasonable for us in the future to look at the constraints on our model not only from the most recent DM direct searches, but also from LHC measurements on the gauge and Yukawa couplings of the $125\,{\rm GeV}$ Higgs boson and on its invisible decay mode, as well as from some upcoming requirements.  

\acknowledgments

The work of PC is financially supported by the Institute for the Promotion of Teaching Science and Technology (IPST) under the project of the \lq\lq Research Fund for DPST Graduate with First Placement\rq\rq\, under Grant No.033/2557 and by the Thailand Research Fund (TRF) under the project of the \lq\lq TRF Grant for New Researcher\rq\rq with Grant No.TRG5780143 and CX is supported by the research funds from the Institute of Advanced Studies and the School of Physical and Mathematical Sciences, Nanyang Technological University, Singapore.


\begin{thebibliography}{99}

\bibitem{Feng:2010gw}
  J.~L.~Feng,
  ``Dark Matter Candidates from Particle Physics and Methods of Detection,''
  Ann.\ Rev.\ Astron.\ Astrophys.\  {\bf 48} (2010) 495
  
\bibitem{Feng:2003uy} 
  J.~L.~Feng, A.~Rajaraman and F.~Takayama, ``SuperWIMP dark matter signals from the early universe,''
  Phys.\ Rev.\ D {\bf 68}, 063504 (2003)
  
\bibitem{Viel:2005qj} 
  M.~Viel, J.~Lesgourgues, M.~G.~Haehnelt, S.~Matarrese and A.~Riotto, ``Constraining warm dark matter candidates including sterile neutrinos and light gravitinos with WMAP and the Lyman-alpha forest,''
  Phys.\ Rev.\ D {\bf 71}, 063534 (2005)
  
  \bibitem{Starobinsky:1979ty} 
  A.~A.~Starobinsky,
 ``Relict Gravitation Radiation Spectrum and Initial State of the Universe. (In Russian),''
  JETP Lett.\  {\bf 30}, 682 (1979)
  [Pisma Zh.\ Eksp.\ Teor.\ Fiz.\  {\bf 30}, 719 (1979)].
  
  \bibitem{Starobinsky:1980te} 
  A.~A.~Starobinsky,
  ``A New Type of Isotropic Cosmological Models Without Singularity,''
  Phys.\ Lett.\ B {\bf 91}, 99 (1980).

  \bibitem{Guth:1980zm} 
  A.~H.~Guth,
  ``The Inflationary Universe: A Possible Solution to the Horizon and Flatness Problems,''
  Phys.\ Rev.\ D {\bf 23}, 347 (1981)
  
  \bibitem{Linde:1981mu} 
  A.~D.~Linde,
  ``A New Inflationary Universe Scenario: A Possible Solution of the Horizon, Flatness, Homogeneity, Isotropy and Primordial Monopole Problems,''
  Phys.\ Lett.\ B {\bf 108}, 389 (1982).
  
  \bibitem{Albrecht:1982wi} 
  A.~Albrecht and P.~J.~Steinhardt,
  ``Cosmology for Grand Unified Theories with Radiatively Induced Symmetry Breaking,''
  Phys.\ Rev.\ Lett.\  {\bf 48}, 1220 (1982).
  
\bibitem{Ade:2015lrj} 
P.~A.~R.~Ade {\it et al.} [Planck Collaboration], ``Planck 2015 results. XX. Constraints on inflation,'' arXiv:1502.02114.


\bibitem{Bezrukov:2007ep}
  F.~L.~Bezrukov and M.~Shaposhnikov,
  ``The Standard Model Higgs boson as the inflaton,'' Phys.\ Lett.\ B {\bf 659} (2008) 703 [arXiv:0710.3755 [hep-th]].
  
\bibitem{Bezrukov:2008ut}
  F.~Bezrukov, D.~Gorbunov and M.~Shaposhnikov,
  ``On initial conditions for the Hot Big Bang,'' JCAP {\bf 0906} (2009) 029 [arXiv:0812.3622 [hep-ph]]
  
\bibitem{Inagaki:1997kz} 
  T.~Inagaki, T.~Muta and S.~D.~Odintsov, ``Dynamical symmetry breaking in curved space-time: Four fermion interactions,''
  Prog.\ Theor.\ Phys.\ Suppl.\  {\bf 127}, 93 (1997)
  
  \bibitem{Clark09} 
  T.~E.~Clark, B.~Liu, S.~T.~Love and T.~ter Veldhuis,
  ``The Standard Model Higgs Boson-Inflaton and Dark Matter,''  Phys.\ Rev.\ D {\bf 80}, 075019 (2009)  [arXiv:0906.5595 [hep-ph]].

\bibitem{Lerner:2009xg} 
  R.~N.~Lerner and J.~McDonald, ``Gauge singlet scalar as inflaton and thermal relic dark matter,''
  Phys.\ Rev.\ D {\bf 80}, 123507 (2009)
  
\bibitem{Cline:2012hg} 
  J.~M.~Cline and K.~Kainulainen, ``Electroweak baryogenesis and dark matter from a singlet Higgs,''
  JCAP {\bf 1301}, 012 (2013)
   
 \bibitem{Channuie16}
 P.~Channuie and C.~Xiong, (in preparation).
  
\bibitem{Nambu:1961tp} 
  Y.~Nambu and G.~Jona-Lasinio, ``Dynamical Model of Elementary Particles Based on an Analogy with Superconductivity. 1.,''
  Phys.\ Rev.\  {\bf 122}, 345 (1961).
  
\bibitem{Nambu:1961fr} 
  Y.~Nambu and G.~Jona-Lasinio, ``Dynamical Model Of Elementary Particles Based On An Analogy With Superconductivity. Ii,''
  Phys.\ Rev.\  {\bf 124}, 246 (1961).
  
\bibitem{Inagaki:2015eza} 
  T.~Inagaki, S.~D.~Odintsov and H.~Sakamoto, ``Gauged Nambu-Jona-Lasinio inflation,''
  Astrophys.\ Space Sci.\  {\bf 360}, no. 2, 67 (2015)
  
\bibitem{Inagaki:2016vkf}
  T.~Inagaki, S.~D.~Odintsov and H.~Sakamoto, ``Inflation from the Finite Scale Gauged Nambu-Jona-Lasinio Model,''
  arXiv:1611.00210 [hep-ph].
  
\bibitem{Xiong:2013uuf} 
  C.~Xiong, ``QCD Flux Tubes and Anomaly Inflow,''
  Phys.\ Rev.\ D {\bf 88}, no. 2, 025042 (2013)
  
\bibitem{Xiong:2014yba}
  C.~Xiong, ``Gauged Nambu-Jona-Lasinio model and axionic QCD string,''
  arXiv:1412.8759 [hep-ph]
  
\bibitem{Channuie:2011rq}
  P.~Channuie, J.~J.~Joergensen and F.~Sannino,
  ``Minimal Composite Inflation,''
  JCAP {\bf 1105} (2011) 007
  
\bibitem{Bezrukov:2011mv}
  F.~Bezrukov, P.~Channuie, J.~J.~Joergensen and F.~Sannino,
  ``Composite Inflation Setup and Glueball Inflation,''
  Phys.\ Rev.\ D {\bf 86} (2012) 063513
  
\bibitem{Channuie:2012bv}
  P.~Channuie, J.~J.~Jorgensen and F.~Sannino,
  ``Composite Inflation from Super Yang-Mills, Orientifold and One-Flavor QCD,''
  Phys.\ Rev.\ D {\bf 86} (2012) 125035
  
\bibitem{Bilenky:2010zza} 
  S.~Bilenky, ``Introduction to the physics of massive and mixed neutrinos,''
  Lect.\ Notes Phys.\  {\bf 817}, 1 (2010).
  
\bibitem{Hill:2002ap} 
  C.~T.~Hill and E.~H.~Simmons, ``Strong dynamics and electroweak symmetry breaking,''
  Phys.\ Rept.\  {\bf 381}, 235 (2003)
  Erratum: [Phys.\ Rept.\  {\bf 390}, 553 (2004)]
  
\bibitem{Martin:1997ns} 
  S.~P.~Martin, ``A Supersymmetry primer,'' Adv.\ Ser.\ Direct.\ High Energy Phys.\  {\bf 21}, 1 (2010)
  [Adv.\ Ser.\ Direct.\ High Energy Phys.\  {\bf 18}, 1 (1998)]
  
\bibitem{Inagaki:2015fva} 
  T.~Inagaki, R.~Nakanishi and S.~D.~Odintsov, ``Non-Minimal Two-Loop Inflation,''
  Phys.\ Lett.\ B {\bf 745}, 105 (2015)

\bibitem{Hatsuda} 
  T.~Hatsuda and T.~Kunihiro,
  ``QCD phenomenology based on a chiral effective Lagrangian,''
  Phys.\ Rept.\  {\bf 247}, 221 (1994).

\bibitem{Klevansky} 
  S.~P.~Klevansky,
  ``The Nambu-Jona-Lasinio model of quantum chromodynamics,''  
  Rev.\ Mod.\ Phys.\  {\bf 64}, 649 (1992).  

\bibitem{NMSM}
  H.~Davoudiasl, R.~Kitano, T.~Li, H.~Murayama,
  ``The New minimal standard model,''
  Phys.\ Lett.\  {\bf B609}, 117-123 (2005).
  [hep-ph/0405097].

\bibitem{Hill91} 
  C.~T.~Hill and D.~S.~Salopek,
  ``Calculable nonminimal coupling of composite scalar bosons to gravity,''  Annals Phys.\  {\bf 213}, 21 (1992).  
  
\bibitem{Geyer:1996kg} 
  B.~Geyer and S.~D.~Odintsov, ``Chiral symmetry breaking in gauged NJL model in curved space-time,''
  Phys.\ Rev.\ D {\bf 53}, 7321 (1996)
  
\bibitem{Buchbinder:1985ba} 
  I.~L.~Buchbinder and S.~D.~Odintsov, ``Asymptotical Conformal Invariance In Curved Space-time,''
  Lett.\ Nuovo Cim.\  {\bf 42}, 379 (1985)
  
\bibitem{Lyth:1998xn}
  D.~H.~Lyth and A.~Riotto,
  ``Particle physics models of inflation and the cosmological density perturbation,'' Phys.\ Rept.\  {\bf 314} (1999) 1 [hep-ph/9807278].
  
\bibitem{Ebert:1997fc} 
  D.~Ebert, ``Bosonization in particle physics,''
  Lect.\ Notes Phys.\  {\bf 508}, 103 (1998)
  
\bibitem{Xiong:2016fum} 
  C.~Xiong, ``A de-gauging approach to physics beyond the Standard Model,'' arXiv:1606.01883 [hep-ph]. {\it Proceedings of the Conference New Physics at the Large Hadron Collider, Singapore, 2016}. 
  
\bibitem{Xiong:2016mxu} 
  C.~Xiong, ``Dark fermions from the Standard Model via spin-charge separation,''
  arXiv:1605.09786 [hep-ph].
  
\bibitem{Cline:2013gha} 
  J.~M.~Cline, K.~Kainulainen, P.~Scott and C.~Weniger, ``Update on scalar singlet dark matter,''
  Phys.\ Rev.\ D {\bf 88}, 055025 (2013) Erratum: [Phys.\ Rev.\ D {\bf 92}, no. 3, 039906 (2015)]
  
\bibitem{Gondolo:1990dk} 
  P.~Gondolo and G.~Gelmini, ``Cosmic abundances of stable particles: Improved analysis,'' Nucl.\ Phys.\ B {\bf 360}, 145 (1991)
    
\bibitem{Lee:1977ua} 
  B.~W.~Lee and S.~Weinberg, ``Cosmological Lower Bound on Heavy Neutrino Masses,''
  Phys.\ Rev.\ Lett.\  {\bf 39}, 165 (1977).
  
\bibitem{Hung:2010xh} 
  P.~Q.~Hung and C.~Xiong,
  ``Dynamical Electroweak Symmetry Breaking with a Heavy Fourth Generation,''
  Nucl.\ Phys.\ B {\bf 848}, 288 (2011)

\bibitem{Hung:2009hy} 
  P.~Q.~Hung and C.~Xiong,
  ``Renormalization Group Fixed Point with a Fourth Generation: Higgs-induced Bound States and Condensates,''
  Nucl.\ Phys.\ B {\bf 847}, 160 (2011)
  
\bibitem{Aravind:2015xst} 
  A.~Aravind, M.~Xiao and J.~H.~Yu, ``Higgs Portal to Inflation and Fermionic Dark Matter,''
  Phys.\ Rev.\ D {\bf 93}, no. 12, 123513 (2016)
 
\bibitem{Barger:2012hv} 
  V.~Barger, M.~Ishida and W.~Y.~Keung, ``Total Width of 125 GeV Higgs Boson,''
  Phys.\ Rev.\ Lett.\  {\bf 108}, 261801 (2012)
 
\bibitem{Campbell:2016llw} 
  R.~Campbell, S.~Godfrey and A.~de la Puente, ``The Dilaton-like Higgs boson with scalar singlet dark matter,''
  arXiv:1607.02158 [hep-ph].
  
\bibitem{Ade:2013zuv} 
  P.~A.~R.~Ade {\it et al.} [Planck Collaboration], ``Planck 2013 results. XVI. Cosmological parameters,''
  Astron.\ Astrophys.\  {\bf 571}, A16 (2014)
  
\bibitem{Guo:2010hq} 
  W.~L.~Guo and Y.~L.~Wu, ``The Real singlet scalar dark matter model,''
  JHEP {\bf 1010}, 083 (2010)
  
\bibitem{Han:2015hda} 
  H.~Han and S.~Zheng, ``New Constraints on Higgs-portal Scalar Dark Matter,''
  JHEP {\bf 1512}, 044 (2015)
  
\bibitem{Channuie:2016xmq} 
  P.~Channuie and P.~Koad, ``Preheating after technicolor inflation,''
  Phys.\ Rev.\ D {\bf 94}, no. 4, 043528 (2016)  
  
\bibitem{Akerib:2016vxi} 
  D.~S.~Akerib {\it et al.}, ``Results from a search for dark matter in LUX with 332 live days of exposure,''
  arXiv:1608.07648 [astro-ph.CO]
  
\bibitem{Tan:2016zwf} 
  A.~Tan {\it et al.} [PandaX-II Collaboration], ``Dark Matter Results from First 98.7-day Data of PandaX-II Experiment,''
  arXiv:1607.07400 [hep-ex]

\bibitem{He:2016} 
X.-G.~He and J.~Tandean, ``New LUX and PandaX-II Results Illuminating the Simplest Higgs-Portal Dark Matter Models,'' arXiv:1609.03551 [hep-ph]

\end{thebibliography}
\end{document}